\documentclass[a4paper,onecolumn,11pt,accepted=2024-12-03]{quantumarticle}
\pdfoutput=1
\DeclareUnicodeCharacter{2212}{-}
\usepackage{hyperref}
\usepackage{doi}
\usepackage{soul}
\usepackage{marginnote}
 \usepackage{graphicx}
 \usepackage{xcolor,cite,etoolbox}
\usepackage{amsmath,amsthm,amsfonts,amscd,amssymb,mathrsfs,cite}
\usepackage[initials]{amsrefs} 
\usepackage{tikz}
\usepackage{amscd}   
\usepackage[all]{xy} 
\usepackage{booktabs} 
\usepackage{stmaryrd}
\usepackage{mathtools}
\usepackage{numprint}
\usepackage{epstopdf}
\usepackage{qcircuit}
\usepackage[utf8]{inputenc} 
\usepackage{soul} 
\usepackage{nccmath} 
\usepackage{url}
\DeclareMathAlphabet{\mathscr}{OT1}{pzc}{m}{it} 

\usepackage{hyperref}
  \hypersetup{colorlinks=true,citecolor=blue, urlcolor= cyan}

 \AtBeginDocument{%
    \def\MR#1{} 
 }

\usepackage{color}

\theoremstyle{remark}

\usepackage{physics}
\renewcommand{\sl}{\mathfrak{sl}}
\usepackage{authblk}

\title{Hexagons govern three-qubit contextuality}

\author{Metod Saniga}
\affiliation{Astronomical Institute of the Slovak Academy of Sciences, SK-05960 Tatransk\' a Lomnica, Slovakia}
\author{Frédéric Holweck}
\affil{Laboratoire Interdisciplinaire Carnot de Bourgogne, ICB/UTBM, UMR 6303, CNRS, Universit\'e de Technologie de Belfort-Montbéliard, F-90010 Belfort Cedex, France}
\affil{Department of Mathematics and Statistics, Auburn University, Auburn, AL, USA}
\author{Colm Kelleher}
\affil{Laboratoire Interdisciplinaire Carnot de Bourgogne, ICB/UTBM, UMR 6303, CNRS, Universit\'e de Technologie de Belfort-Montbéliard, F-90010 Belfort Cedex, France}
\author{Axel Muller}
\affil{Université Marie et Louis Pasteur, CNRS, Institut FEMTO-ST, F-25000 Besançon, France}
\author{Alain Giorgetti}
\affil{Université Marie et Louis Pasteur, CNRS, Institut FEMTO-ST, F-25000 Besançon, France}
\author{Henri de Boutray}
\affil{ColibriTD, F-75013 Paris, France}

\date{}

\begin{document}

\maketitle

\abstract{

Split Cayley hexagons of order two are distinguished finite geometries living in the three-qubit symplectic polar space in two different forms, called classical and skew. Although neither of the two yields observable-based contextual configurations of their own, {\it classically}-embedded copies are found to fully encode contextuality properties of the most prominent three-qubit contextual configurations in the following sense: for each set of unsatisfiable contexts of such a contextual configuration
there exists some classically-embedded hexagon sharing with the configuration
exactly this set of contexts and nothing else.  We demonstrate this fascinating property first on the configuration comprising all 315 contexts of the space and then on doilies,
both types of quadrics as well as on complements of skew-embedded hexagons.
In connection with the last-mentioned case and elliptic quadrics we also conducted some experimental tests on a Noisy Intermediate Scale Quantum (NISQ) computer to substantiate our theoretical findings.

 }

\section{Introduction}
Finite geometry is a branch of mathematics that deals with geometries made of a finite number of points, lines and/or linear spaces of higher dimensions. This perspective of working with spaces that contain only a finite number of geometric elements is rather counter-intuitive and far from the intuition provided, for example, by Euclidean geometry as it lacks concepts like smoothness, differentiability, distance {\it etc}. Gino Fano~\cite{fano} was one of the first geometers to formalize this idea and his name is now associated with the smallest finite projective plane, i.\,e. the Fano plane (Figure~\ref{fig:fano}).
\begin{figure}[!ht]
\begin{center}
\begin{tikzpicture}
\tikzstyle{point}=[ball color=black, circle, draw=black, inner sep=0.1cm]
\node (v7) at (0,0) [point] {};
\draw (0,0) circle (1cm);
\node (v1) at (90:2cm) [point] {};
\node (v2) at (210:2cm) [point] {};
\node (v4) at (330:2cm) [point] {};
\node (v3) at (150:1cm) [point] {};
\node (v6) at (270:1cm) [point] {};
\node (v5) at (30:1cm) [point] {};
\draw (v1) -- (v3) -- (v2);
\draw (v2) -- (v6) -- (v4);
\draw (v4) -- (v5) -- (v1);
\draw (v3) -- (v7) -- (v4);
\draw (v5) -- (v7) -- (v2);
\draw (v6) -- (v7) -- (v1);
\end{tikzpicture}
\caption{The Fano plane depicted in its standard rendering. This self-dual geometry comprises seven points (bullets) and seven lines (six straight segments and a circle), with three points per line and, dually, three lines through a point, being equivalent to the projective plane over the two-element field $\mathbb{F}_2=\{0,1\}$.}\label{fig:fano}
\end{center}
\end{figure}

Over the past 20 years (see, for example, \cites{sp06,hs08,slp12,lps13,ls17,koen09} and/or \cites{wae14,waeara13} for a slightly different, more heuristic approach), finite geometry has been introduced into the field of quantum information and mathematical physics to mainly model and analyse the commutation relations within the $n$-qubit Pauli group, $\mathcal{P}_n$, defined by
\begin{equation}
    \mathcal{P}_n=\{sA_1A_2\dots A_n: A_i\in \{I,X,Y,Z\}, s\in\{\pm 1,\pm i\}\},
    \label{Pauli}
\end{equation}
where $X$, $Y$ and $Z$ are the famous Pauli matrices, $I$ is the associated $2 \times 2$ identity matrix and
\begin{equation}
A_1 A_2\dots A_n \equiv A_1\otimes A_2\otimes \dots \otimes A_n. 
 \label{tp}
\end{equation}
For example, for $n=3$, if one ignores the global phase of each operator, a maximum set of mutually commuting three-qubit Pauli  operators (disregarding the identity) forms a Fano plane, as shown with the example provided by Figure~\ref{fig:fano-observable}.  Two points are
collinear if and only if the matrix product of the observables labelling them commutes.

\begin{figure}[!ht]
\begin{center}
\begin{tikzpicture}
\tikzstyle{point}=[ball color=black, circle, draw=black, inner sep=0.1cm]
\node (v7) at (0,0) [point] {};
\draw (0,0) circle (1cm);
\node (v1) at (90:2cm) [point] {};
\node (v2) at (210:2cm) [point] {};
\node (v4) at (330:2cm) [point] {};
\node (v3) at (150:1cm) [point] {};
\node (v6) at (270:1cm) [point] {};
\node (v5) at (30:1cm) [point] {};
\node (v11) at (90:2.35cm)  {$YXY$};
\node (v12) at (210:2.7cm)  {$XXX$};
\node (v14) at (330:2.7cm) {$YYX$};
\node (v13) at (150:1.55cm) {$ZIZ$};
\node (v16) at (270:1.35cm) {$ZZI$};
\node (v15) at (30:1.55cm)  {$IZZ$};
\node (v17) at (0.65,0) {$~XYY$};
\draw (v1) -- (v3) -- (v2);
\draw (v2) -- (v6) -- (v4);
\draw (v4) -- (v5) -- (v1);
\draw (v3) -- (v7) -- (v4);
\draw (v5) -- (v7) -- (v2);
\draw (v6) -- (v7) -- (v1);
\end{tikzpicture}
\end{center}
\caption{The Fano plane labelled by three-qubit observables encapsulates the fact that these seven observables form a set of mutually commuting operators. In fact, it is an example of a maximal set of mutually commuting observables in the three-qubit Pauli group.}\label{fig:fano-observable}
\end{figure}

In what follows, a positive or negative \emph{(quantum) context} 
will be a set of mutually commuting observables such that their product is $+I^{\otimes n}$ or $-I^{\otimes n}$, respectively, where
$I^{\otimes n} \equiv I_{(1)} \otimes I_{(2)} \otimes \dots \otimes I_{(n)}$ is the $n$-fold tensor product of $2 \times 2$ identity matrices. For instance, the seven observables of Figure~\ref{fig:fano-observable} form a (negative) three-qubit context. A contextual configuration will be an arrangement of observables made of contexts such that there is no Non-Contextual Hidden Variable (NCHV) model that can reproduce the outcomes predicted by the rules of Quantum Mechanics (QM).
\begin{figure}[!ht]
\begin{center}
\includegraphics[width=10cm]{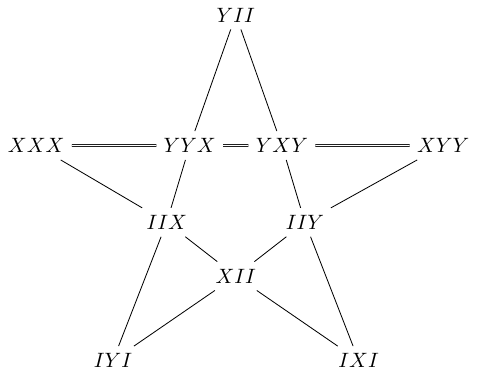}
\end{center}
\caption{Mermin pentagram: This configuration of ten three-qubit observables provides an operator-based proof of the Kochen-Specker Theorem. There is no NCHV model that can reproduce the outcomes predicted by QM for this configuration.}\label{fig:mermin-pentagram}
\end{figure}

Let us consider, for example, the configuration portrayed in Figure~\ref{fig:mermin-pentagram}, known as the Mermin pentagram.
Each node of the configuration is a three-qubit Pauli operator whose eigenvalues are
$+1$ and $-1$. Each line of the configuration represents a context. The doubled line is the unique context with a negative sign, i.\,e. the product of the observables on this context is $-I^{\otimes 3}$. David Mermin introduced this configuration in~\cite{mermin} as an alternative proof of the famous Kochen-Specker (KS) Theorem~ (see, for example, \cites{kosp67,appl05,speck60,bell66,budroni2022kochen}). 
Recall that the Kochen-Specker Theorem is a no-go result that proves the non-existence of an NCHV model, i.\,e. it proves that a Hidden Variable (HV) model that would reproduce the outcomes predicted by QM has to be context-dependent. Consider our Mermin pentagram. An HV model that reproduces predictions of QM for this set of operators should satisfy the following constraints:
\begin{enumerate}
    \item Each node gets assigned a pre-definite measurement value $\pm 1$.
    \item The product of the measurements on each context (line) should be of the same sign as the context itself.
\end{enumerate}
The second constraint stems from the fact that the product of the eigenvalues of a set of mutually commuting observables should be an eigenvalue of the product of the observables. As it can readily be discerned from Figure~\ref{fig:mermin-pentagram}, this sign constraint is not satisfiable unless the pre-definite values on the nodes are context-dependent. Note that the negative line of the Mermin pentagram of Figure~\ref{fig:mermin-pentagram} is nothing but the set of four mutually commuting observables one obtains from the Fano plane provided by Figure~\ref{fig:fano-observable} by removing the (circled) line and its points.  

Observable-based proof of the KS Theorem can already be built from two-qubit Pauli observables, like in the Mermin-Peres magic square, see Section~\ref{sec:degree} and~\cites{peres1991two,mermin}. For more information on the two-qubit case, the reader is referred to consult~\cites{DHGMS22,HS17}, as well as~\cite{budroni2022kochen} for a recent survey on contextuality.    

In this paper we will focus solely on the three-qubit case ($n = 3$) to demonstrate in detail how a specific arrangement of three-qubit contexts isomorphic to the smallest split Cayley hexagon underpins contextuality properties of a whole class of distinguished aggregates of three-qubit observables. This hexagon made its debut in physics as early as 2008 in~\cite{lsv}, where the authors employed its particular three-qubit subgeometry to model the $E_7$-symmetric black-hole entropy formula in string theory. Later~\cite{psh}, it was pointed out that the order of the automorphism group of the hexagon, \numprint{12096}, coincides with the total number of three-qubit Mermin pentagrams. 
At about the same time~\cite{sppl}, it was already noticed  that some contextual three-qubit configurations can be uniquely extended to geometric hyperplanes of the hexagon.

It was however only very recently~\cite{holweck2022three} that the fact that the hexagon embeds in the three-qubit space in two different ways has been properly taken into account and recognized as having deep physical meaning, leading to our present study.

The paper is organized as follows. In Section~\ref{sec:degree} we introduce the notion of the symplectic polar space of rank $n$ and order two, $\mathcal{W}(2n-1,2)$, which is the geometric framework for the commutation relations within the $n$-qubit generalized Pauli group. Next, we define the most prominent subgeometries of this space, quadrics, and show that the particular space $\mathcal{W}(5,2)$ for three-qubits is endowed with another remarkable kind of subgeometry, namely the split Cayley hexagon of order two, which occurs in this space in two non-isomorphic embeddings. Then we introduce the notion of the degree of contextuality of a quantum configuration, that measures how far a contextual configuration of observables is from supporting an NCHV model, and explicitly illustrate
this notion on the example of the smallest symplectic space $\mathcal{W}(3,2)$, the doily, and one of its hyperbolic quadrics.
Section~\ref{sec:skew} deals with the two types of hexagon's embeddings. We first discuss the principal difference between them making use of sets of nine mutually disjoint planes of $\mathcal{W}(5,2)$. 
Then we show how the intersections of a skew-embedded hexagon with various doilies (i.\,e. $\mathcal{W}(3,2)$'s) help us to better understand the fact that the complement of this hexagon is a contextual configuration.
Section~\ref{sec:W52}, the core section of the paper, first highlights the  facts that the degree of contextuality of the configuration comprising all 315 line-contexts of $\mathcal{W}(5,2)$ is equal to $63$ and that the corresponding 63 unsatisfiable contexts of this configuration are the 63 lines of a copy of the split Cayley hexagon of order two that is embedded {\it classically} into $\mathcal{W}(5,2)$. The first property is then substantiated by a chain of group-theoretic and algebro-geometric arguments. 
After introducing a line-layered decomposition of the hexagon and describing how this decomposition helps us to quickly alternate between the two embeddings, we illustrate on several examples our most crucial finding, namely that contextuality properties of most prominent families of contextual three-qubit configurations having three-element contexts, entailing doilies, elliptic quadrics and hyperbolic quadrics, are fully described in terms of their intersection with properly-selected classically-embedded copies of the hexagon.  
As a substantiation of our findings, Section~\ref{sec:ibm} outlines the results of testing a specific contextual inequality introduced by Cabello on the IBM Quantum Experience by employing an elliptic quadric and the complement of a particular copy of skew-embedded hexagon of $\mathcal{W}(5,2)$. Our procedure follows and improves that of~\cite{holweck2021testing}, where the contextuality of $\mathcal{W}(5,2)$ as a whole was already successfully tested. 
Finally, Section~\ref{sec:conclusion} summarizes the main achievements, makes some proposals of how to tackle the next case, that of four qubits, in a similar unifying way  and briefly addresses an intriguing formal analogy between the {\it two} inequivalent embeddings of the smallest split Cayley hexagon into the three-qubit $\mathcal{W}(5,2)$ and {\it two} inequivalent kinds of genuine tripartite entanglement.

\section{Contextual configurations in symplectic polar spaces, degree of contextuality}\label{sec:degree}

Over the past $15$ years symplectic geometry over the two-element field has been investigated in the context of quantum information as it encodes the commutation relations between the elements of the $n$-qubit Pauli group~\cites{saniga2007multiple,havlicek2009factor}. 
Recall that the Pauli matrices $X,Y,Z$ and the identity operator $I$ can be given in terms of $X$ and $Z$ as follows (phase omitted): $X=Z^0.X, Y=Z^1.X^1, Z=Z^1.X^0$ and $I=Z^0.X^0$. Consider the map:
\begin{equation}
\begin{array}{ccc}
\mathcal{P}_n/\{\pm I^{\otimes n},\pm iI^{\otimes n}\} & \to & \mathbb{F}_2^{2n}\\
(Z^{a_1}.X^{b_1})\otimes (Z^{a_2}.X^{b_2})\otimes \dots \otimes (Z^{a_n}.X^{b_n}) & \mapsto & (a_1,a_2,\dots,a_n,b_1,b_2,\dots,b_n).
\end{array}
\label{Pauli-map}
\end{equation}

This map associates bijectively to any $n$-qubit Pauli matrix (up to a phase) a unique vector of $\mathbb{F}_2^{2n}$. Thinking projectively, one can associate to any (phase disregarded) non-trivial operator of $\mathcal{P}_n/\{\pm I^{\otimes n},\pm iI^{\otimes n}\}$ a unique point of PG$(2n-1,2)$, the $(2n-1)$-dimensional projective space over the two-element field\footnote{Let $V$ be a ($d+1$)-dimensional vector space over $\mathbb{F}_2$. The projective space PG$(d,2)$ is the geometry whose points, lines, planes, etc. are the vector subspaces of $V$ of respective dimensions 1, 2, 3, etc.. Its points can be represented by ($d+1$)-tuples of the form $(x_1, x_2, x_3,\ldots, x_{d+1})$ where $x_i \in \mathbb{F}_2$, disregarding the trivial $(0,0,0,\ldots, 0)$-tuple.}.

Let us equip the projective space PG$(2n-1,2)$ with the symplectic form defined as
(note that $+1 =-1$ over $\mathbb{F}_2$)
\begin{equation}
\sigma(x,y)=x_1y_{n+1}+x_2y_{n+2}+\dots +x_n y_{2n}+x_{n+1}y_1+x_{n+2}y_2+\dots +x_{2n}y_n
\label{polarity}
\end{equation}
and call a subspace of PG$(2n-1,2)$ \emph{totally isotropic} if this form vanishes identically on it, i.\,e. $\sigma(x,y) = 0$ for any two distinct
points $x$ and $y$ of the subspace.
Then the space $\mathcal{W}(2n-1,2)$ of all totally isotropic subspaces of PG$(2n-1,2)$ with respect to $\sigma$ is called the symplectic polar space of rank $n$ and order $2$. This space encodes the commutation relations of $\mathcal{P}_n$ in the sense that one can check by a straightforward calculation that any pair of collinear points $x,y\in \mathcal{W}(2n-1,2)$ corresponds to two commuting observables $O_x$ and $O_y$ in $\mathcal{P}_n$, see e.\,g. \cite{havlicek2009factor}. 
In what follows, it will be implicitly assumed that all the points of such multi-qubit $\mathcal{W}(2n-1,2)$ are labelled solely by canonical observables of $\mathcal{P}_n$, i.\,e. by those observables whose phase $s$ equals $1$
(see eq.\,(\ref{Pauli})). Moreover, when referring to a subspace of $\mathcal{W}(2n-1,2)$ we will always have in mind also the associated set of pairwise commuting observables, and {\it vice versa}.

A large number of contextual configurations can thus be advantageously identified with {\it subgeometries} of $\mathcal{W}(2n-1,2)$. Indeed, as contexts are sets of mutually commuting observables whose products are, up to a sign, equal to the identity operator, then all lines, planes, and more generally linear subspaces of $\mathcal{W}(2n-1,2)$ can be chosen as contexts. For example, the Fano plane depicted in Figure~\ref{fig:fano-observable} is a linear subspace of the largest dimension in $\mathcal{W}(5,2)$.
The most prominent subgeometries of $\mathcal{W}(2n-1,2)$ are quadrics, which occur in two different forms.  A {\it hyperbolic} quadric $\mathcal{Q}^{+}(2n - 1, 2)$, $n \geq 1$, is a subgeometry whose
equation can be brought to the following standard form
\begin{equation}
x_1 x_{n+1} + x_2 x_{n+2} \ldots + x_{n} x_{2n} = 0.
\label{hqsteqn}
\end{equation}
Each $\mathcal{Q}^{+}(2n - 1, 2)$ contains 
\begin{equation}
|\mathcal{Q}^{+}|_p = (2^{n-1} + 1)(2^{n} -1) 
\label{ptsonhq}
\end{equation}
points, 
\begin{equation}
|\mathcal{Q}^{+}|_l = \frac{1}{3}(2^n - 1)(2^{n-2} + 1)(2^{2(n-1)} - 1) 
\label{lnsonhq}
\end{equation}
lines and there are 
\begin{equation}
|W|_{\mathcal{Q}^{+}} = |\mathcal{Q}^{+}|_p + 1 = (2^{n-1} + 1)(2^{n} -1) + 1
\label{hqinwn}
\end{equation}
copies of them in $\mathcal{W}(2n-1,2)$.
An {\it elliptic} quadric $\mathcal{Q}^{-}(2n - 1,2)$, $n \geq 2$,
comprises all points and subspaces of $\mathcal{W}(2n-1,2)$ satisfying the standard equation 
\begin{equation}
f(x_1,x_{n+1})+x_2 x_{n+2}+\cdots+x_{n}x_{2n} = 0, 
\label{eqsteqn}
\end{equation}
where $f$ is an irreducible polynomial over $\mathbb{F}_2$.
Each $\mathcal{Q}^{-}(2n - 1, 2)$ contains 
\begin{equation}
|\mathcal{Q}^{-}|_p = (2^{n-1} - 1)(2^{n} + 1) 
\label{ptsoneq}
\end{equation}
points, 
\begin{equation}
|\mathcal{Q}^{-}|_l = \frac{1}{3}(2^n + 1)(2^{n-2} - 1)(2^{2(n-1)} - 1) 
\label{lnsoneq}
\end{equation}
lines and $\mathcal{W}(2n-1,2)$ features 
\begin{equation}
|W|_{\mathcal{Q}^{-}} = |\mathcal{Q}^{-}|_p + 1 = (2^{n-1} - 1)(2^{n} + 1) + 1
\label{eqinwn}
\end{equation}
copies of them.
Employing the fact that a canonical observable $O$ is either \emph{symmetric} ($O^{{\rm T}} = O$) or \emph{skew-symmetric} ($O^{{\rm T}} = -O$), there exists a quite straightforward way to find all the observables
belonging to a particular quadric without even making use of its algebraic equation and projective coordinates. Namely, it can be readily verified (see, e.\,g., \cite{VL10}) that given a canonical observable $O$,  the set of symmetric canonical observables commuting with $O$ together with the set of skew-symmetric observables not
commuting with $O$ lie on a quadric of  $\mathcal{W}(2N-1,2)$, this quadric being hyperbolic (resp. elliptic) if $O$ is symmetric (resp. skew-symmetric); the observable $O$ is usually called the index of the quadric. An observable is symmetric if the corresponding tensor product (see eq.\,(\ref{tp})) features an even (including zero) number of $Y$'s; otherwise it is skew-symmetric. Also, in order to check whether two different $n$-qubit observables commute it is not necessary to check the (two-way) product
of the corresponding $2^n \times 2^n$ matrices. It suffices to simply count the number of places in which they feature different Pauli matrices; if this number is even the observables commute, if it is odd they do not. Similarly, in order to quickly find the sign of a context one simply takes the bitwise products of the corresponding Pauli matrices and the identity matrix and then multiplies the phases obtained; for example, the three-qubit line consisting of the  
observables $XYZ$, $ZIX$ and $YYY$ is positive as $XYZ . ZIX . YYY$ = $(X.Z.Y)(Y.I.Y)(Z.X.Y)$=$(-iI)(+I)(+iI)$=$+I^{\otimes 3}$.

When it comes to $\mathcal{W}(5,2)$, here we find a particularly remarkable kind of subgeometry, namely the split Cayley hexagon of order two, $\mathcal{H}$, which will be the central object of our paper. 
It is a finite geometry having 63 points and 63 lines, with three points on a line and three lines through a point, such that its smallest polygons are hexagons (hence its name).
To introduce its abstract definition, let us consider the parabolic quadric $\mathcal{Q}$ in the six-dimensional projective space PG$(6,2)$ defined by the following quadratic form
\begin{equation}
Q(x) = x_1 x_4 + x_2 x_5 + x_3 x_6 + x_{7}^{2}.
\label{parquad}
\end{equation}
Then (see, e.\,g., \cite{mal}) the 63 points of $\mathcal{Q}$ and those 63 lines of $\mathcal{Q}$ whose Grassmannian coordinates $p_{ij} \equiv
x_i y_j - x_j y_i$ satisfy the following equations
\begin{equation}
p_{62} = p_{17}, p_{13} = p_{72}, p_{24} = p_{37}, p_{35} = p_{74}, p_{46} = p_{57}, p_{51} = p_{76}, 
p_{14} + p_{25} + p_{36} = 0
\label{grass}
\end{equation}
define a point-line incidence structure isomorphic to  $\mathcal{H}$. In fact, this description of $\mathcal{H}$
is a particular type of the embedding of $\mathcal{H}$ into $\mathcal{Q}$, called {\it classical}. 
There exists, however, another type of embedding of  $\mathcal{H}$ into $\mathcal{Q}$, discovered by Coolsaet~\cite{cool} and referred by him to as {\it skew}, which
is furnished  by the coordinate map 
\begin{equation}
\varepsilon: (x_1, x_2, \ldots, x_7) \mapsto (x_1 + x_6 + f_5(x), x_2 + x_3 + f_4(x), x_3, x_4, x_5, x_6, x_7) 
\label{skew}
\end{equation}
where
\begin{equation}
f_4(x) \equiv x_3 x_5 + x_7 x_4~~~~\text{and}~~~~f_5(x) \equiv x_4 x_6 + x_7 x_5.
\label{fs}
\end{equation}
All the points of a classically-embedded $\mathcal{H}$ are on the same footing. This is, however, not the case with a skew-embedded
 $\mathcal{H}$, where three points that behave `classically' have a special footing; these points lie on a line, called the {\it axis} by Coolsaet. As we work over $\mathbb{F}_2$, we can project $\mathcal{Q}$ into $\mathcal{W}(5,2)$ using the
inverse operation to 
\begin{equation}
(x_1, x_2, \ldots, x_6) \mapsto (x_1, x_2, \ldots, x_6, \sqrt{x_1x_2 + x_3x_4 + x_5x_6})
\label{map}
\end{equation}
to obtain the two (types of) embeddings of $\mathcal{H}$ in $\mathcal{W}(5,2)$.
Similarly to quadrics, also in this case we can completely avoid working with the above-given abstract definitions.
There already exists~\cites{schr,psm} very appealing, and also highly symmetric, graphical visualisations/drawings of the smallest split Cayley hexagon that illustrate all its essential geometric properties and exhibit all its 63 points and 63 lines in a particularly handy way to work with. So, the exposition of our ideas in Sections~\ref{sec:skew} and~\ref{sec:W52} will almost exclusively rest  on such drawings, where
the points of the hexagon are labeled by canonical three-qubit observables in conformity with the
mapping defined by (\ref{Pauli-map}). 
As each hexagon contains all 63 points of $\mathcal{W}(5,2)$, the embedding of a given copy
of the hexagon
into this space is encoded solely in the way how its points are `arranged' into lines of $\mathcal{W}(5,2)$, which in a diagrammatic language translates to a specific way of how the points
of the underlying figure are labelled  by the 63 three-qubit
observables. This also means that one can pass from one copy of the hexagon
to the other by a simple permutation of these three-qubit
labels on the figure in question, the only constraint to be secured during such a procedure is that the three observables associated with any line mutually commute and their (ordinary) product is proportional to $I^{\otimes 3}$.
We will further see that a special kind of permutation that transforms a classically-embedded hexagon to a skew-embedded one, and {\it vice versa}, is intricately related to a line-related
layering of the hexagon, this layering being also essential in understanding the (pivotal) role played by the axis of a skew-embedded copy under such a transformation.  
Also, we shall explicitly illustrate the intersection of selected representatives of the two kinds of quadrics with a classically-embedded hexagon to demonstrate its most baffling contextuality-related role. And all that in a way to be also accessible to the reader having only a very limited background in finite geometry.

The notion of degree of contextuality was introduced in~\cite{DHGMS22} in order to better understand contextual configurations. The \emph{degree of contextuality} of a configuration of observables is the minimal number of contexts that cannot be satisfied by any NCHV model. This degree is the Hamming distance between the image of the incidence matrix of the point-line incidence structure defining the configuration and the vector encoding the signs of the contexts~\cite{DHGMS22}. 
Let $(O,C)$ be a quantum contextual configuration with $p = |O|$ observables
$O=\{M_1,\ldots,M_p\}$ and $l = |C|$ contexts $C=\{c_1,\ldots,c_l\}$.  Its
\emph{incidence matrix} $A \in \mathbb{F}_2^{l \times p}$ is defined by $A_{i,j}
= 1$ if the $i$-th context $c_i$ contains the $j$-th observable $M_j$.
Otherwise, $A_{i,j} = 0$. Its \emph{valuation vector} $E \in \mathbb{F}_2^{l}$
is defined by $E_i = 0$ if $e(c_i) = 1$ and $E_i = 1$ if $e(c_i) = -1$, where
$e$ is the \emph{context valuation} of $(O,C)$ i.\,e. $e(c) = 1$ if the context
$c$ is positive and $e(c) = -1$ if it is negative. Then the degree $d$ of contextuality of $(O,C)$ is defined as follows
 \begin{equation}
  d=d_H(E,\text{Im}(A)),
 \end{equation}
where $d_H$ is the Hamming distance on the vector space $\mathbb{F}_2^l$.
In order to find this degree we proceed as follows. Given a quantum configuration ${K} = (O,C)$, one associates with it a configuration $\widetilde{{K}}$ that is geometrically identical with ${K}$, but has all its observables replaced by $+1$'s and $-1$'s  and the sign of each context $c \in C$ replaced by the product of these $+1$'s and $-1$'s over its members. Then we perform an exhaustive, computer-aided search -- which basically consists of reshuffling and/or swapping these numbers following specific algorithms described in detail in~\cites{muller2023new,muller-4to6qubit}  -- to
find such a $\widetilde{{K}}$ that has the {\it maximum} possible numbers of lines having the same sign as the corresponding contexts in ${K}$. The degree of contextuality of ${K}$ is then equal to the number of remaining lines of $\widetilde{{K}}$, i.\,e. those lines each of which has different parity than its corresponding context in ${K}$; these contexts of ${K}$ that have different signs than the corresponding lines of $\widetilde{{K}}$ are called \emph{unsatisfiable}.

\begin{figure}[t]
\begin{center}
\begin{tikzpicture}[scale=2]
\node (A) at (0,2) {$IZ$};
\node (B) at (1,2) {$ZI$};
\node (C) at (2,2) {$ZZ$};
\node (D) at (0,1) {$XI$};
\node (E) at (1,1) {$IX$};
\node (F) at (2,1) {$XX$};
\node (G) at (0,0) {$XZ$};
\node (H) at (1,0) {$ZX$};
\node (I) at (2,0) {$YY$};
\draw[-]
(A) edge (B)
(A) edge (D)
(B) edge (C)
(B) edge (E)
(C) edge [style=double,thick] (F)
(D) edge (E)
(D) edge (G)
(E) edge (F)
(E) edge (H)
(F) edge [style=double,thick] (I)
(G) edge (H)
(H) edge (I);
\end{tikzpicture}
\begin{tikzpicture}
\tikzstyle{point}=[ball color=black, circle, draw=black, inner sep=0.1cm]
\foreach \x in {18,90,...,306}{
\node  (t\x) at (\x:2.65){};
}

\foreach \x in {54,126,...,342}{
\draw [color=black](\x:1cm) circle (1.17557cm);
}
\draw [style=double, thick](54:1cm) circle (1.17557);
\draw [style=double, thick](126:1cm) circle (1.17557);
\draw [color=black](270:1cm) circle (1.17557);

\fill [white] (0,0) circle (1cm);
\tikzstyle{point}=[color=black, circle, fill, draw=black, inner sep=0.10cm]
\foreach \x in {54,126,...,342}{
\node [point] (o\x) at (\x:2.17){};
\node [point] (i\x) at (\x:1cm) {};
}

\node  (o90) at (90:3){$IY$};
\node  (o18) at (18:3.1){$ZI$};
\node (o54) at (54:2.57) {$ZY$};
\node  (o126) at (126:2.57) {$XY$};
\node (i126) at (150:1.25) {$YX$};
\node  (o162) at (162:3.1){$XI$};
\node (o306) at (306:3.1) {$ZZ$};
\node (o342) at (342:2.57) {$IZ$};
\node (i342) at (350:1.55) {$XZ$};

\node (o270) at (270:2.57) {$YY$};
\node (i270) at (290:1.2) {$YI$};
\node (o234) at (234:3.1) {$XX$};
\node (i198) at (210:1.3) {$ZX$};
\node (o198) at (198:2.6) {$IX$};
\node (i54) at (65:1.3) {$YZ$};

\tikzstyle{point}=[ball color=black, circle, draw=black, inner sep=0.1cm]
\foreach \x in {54,126}{
\node [point] (o\x) at (\x:2.17){};
\node[point] (i\x) at (\x:1cm) {};
}
\tikzstyle{point}=[ball color=black, circle, draw=black, inner sep=0.1cm]
\foreach \x in {90}{
\node [point] (o\x) at (\x:2.65){};}
\node [point] (i270) at (270:1cm) {};
\fill [white] (0,0) circle (1cm);
\foreach \x in {54,126,...,342}{
\node[point] (i\x) at (\x:1cm) {};
\node[point] (o\x) at (\x:2.17557cm) {};
}
 \tikzstyle{point}=[ball color=black, circle, draw=black, inner sep=0.1cm]
\foreach \x in {18,90,...,306}{
\node [point] (t\x) at (\x:2.65){};
} 

\draw [color=black] (t90)--(o126)--(t162)--(o198)--(t234)--(o270)--(t306)--(o342)--(t18)--(o54)--(t90);
\draw (t90)--(i270)--(o270);
\draw (t162)--(i342)--(o342);
\draw (t234)--(i54)--(o54);
\draw (t306)--(i126)--(o126);
\draw (t18)--(i198)--(o198);
\draw [style=double,thick] (t234)--(o270)--(t306);
\draw  (t234)--(o198)--(t162);
\draw  (t234)--(o198)--(t162);
\draw  (o198)--(i198)--(t18);
\draw  (t162)--(i342)--(o342);
\end{tikzpicture}
\end{center}
\caption{A Mermin-Peres magic square, {\it aka} $\mathcal{Q}^{+}(3,2)$, and the doily, {\it aka} $\mathcal{W}(3,2)$, are two `extremal' two-qubit observable-based contextual configurations.  The minimal number of constraints that cannot be satisfied by an NCHV model, i.\,e. the degree of contextuality, is one for the grid and three for the doily. For these `peculiar' configurations, with a single exception, the unsatisfiable constraints can be identified solely with the corresponding negative contexts (represented here by the doubled lines).}\label{fig:mermin-doily}
\end{figure}

For example, if one considers the Magic Mermin pentagram in Figure~\ref{fig:mermin-pentagram}, its degree of contextuality is $1$. 
In the two-qubit case, the Mermin-Peres magic square and the configuration comprising all 15 three-element contexts of two-qubit Pauli matrices, aka $\mathcal{W}(3,2)$, are both contextual with their degree of contextuality being, respectively, $1$ and $3$. Figure~\ref{fig:mermin-doily} illustrates how these configurations can be parametrized by two-qubit observables. The lines/contexts with a negative sign are doubled. It is clear that assigning $+1$ to each node provides a classical model that satisfies all conditions except those imposed by the negative lines. The fact that the degree of contextuality is $1$ and $3$ indicates that one cannot do better with any NCHV model. Note that the grid is a subgeometry of $\mathcal{W}(3,2)$, being isomorphic to $\mathcal{Q}^{+}(3,2)$; it can be shown that there are in fact ten copies of the grid, i.\,e. the triangle-free configuration with nine points and six lines, with three points on a line and two lines through a point, lying in $\mathcal{W}(3,2)$. These two types of contextual two-qubit configurations have two remarkable properties that, as we will see, are absent in the three-qubit (and very likely in any higher rank) case. The first one is the fact that the unsatisfiable contexts do {\it not} cover all the observables of the configuration. 
The other notable fact is that, save for a single copy
of the grid that features three negative contexts, the corresponding unsatisfiable contexts can be identified in each configuration with its negative contexts.

The above-discussed two two-qubit observable-based proofs of the KS Theorem can be considered, together with the Mermin pentagram, as geometrical building blocks of contextuality. Indeed, as recently proved by two of us~\cite{mg23}, once a configuration has been found to be contextual for a particular labeling by Pauli observables, then one can deduce that the same geometric configuration will be contextual and will have the same degree of contextuality no matter what admissible multi-qubit Pauli parametrization is employed. In particular, the fact that the grids and $\mathcal{W}(3,2)$ are contextual implies that $\mathcal{W}(2n-1,2)$ is contextual for all $n\geq 2$ because the symplectic polar space of a given dimension  always contains copies of symplectic polar spaces of a smaller dimension.

\section{The two types of symplectic embeddings of the smallest split Cayley hexagon}\label{sec:skew}

Although the two-qubit symplectic space $\mathcal{W}(3,2)$ already contains the fundamental building blocks furnishing observable-based proofs of quantum contextuality, namely the above-discussed grids as well as two-spreads~\cite{muller2023new}, it is the three-qubit space $\mathcal{W}(5,2)$ where the power of our formalism acquires a completely new dimension. This is mainly because, as already briefly described in Section~\ref{sec:degree}, this space, in addition to quadrics, features also another distinguished subgeometry -- the split Cayley hexagon of order two, $\mathcal{H}$, and in its two
non-isomorphic embeddings at that. 
Recently three of us~\cite{holweck2022three} discovered that the two inequivalent embeddings of this hexagon into $\mathcal{W}(5,2)$ behave differently when it comes to their line-{\it complements}; in particular, it was demonstrated that the complement of any skew-embedded copy of $\mathcal{H}$ is contextual, which is, strangely, not the case for any classically-embedded one. In what follows we will demonstrate that classically-embedded copies of $\mathcal{H}$ also enter the game, but in a different and rather
unexpected way. To this end in view, it is necessary to address first in more detail the principal difference between the two embeddings.

\subsection{Classical versus skew embeddings of the hexagon}

To see the fundamental difference between the two embeddings, let us call a point of the split Cayley hexagon of order two located in $\mathcal{W}(5,2)$ 
{\it planar} if all the three lines passing through it lie in a plane of $\mathcal{W}(5,2)$ i.\,e. the seven observables lying on these three lines mutually commute. 
A classically-embedded hexagon enjoys the property that {\it each} of its points is planar. In a skew-embedded hexagon, however, there are only 15 points
that are planar; they lie on six lines forming three concurrent pairs, the three points of concurrence lying on the {\it axis} of the hexagon. 
For each of the remaining 48 points only two lines passing through it lie in a plane of $\mathcal{W}(5,2)$.

\begin{figure}[t]
\vspace*{-0.cm}
\centerline{\includegraphics[width=7cm,clip=]{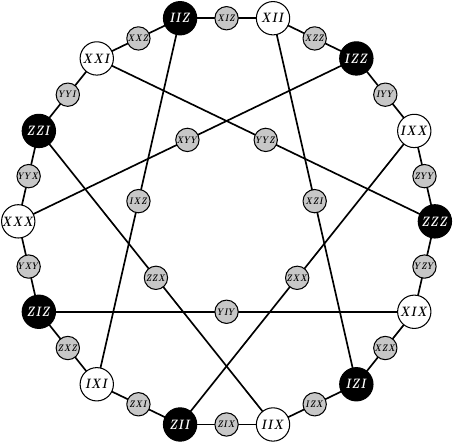}}
\vspace*{-0.0cm}
\caption{A copy of the Heawood graph accommodating, as sets of observables, a pair of three-qubit Fano planes, one represented by seven black bullets and the other by seven big circles; also shown are the remaining 21 points (gray) on the lines represented by the 21 edges of the graph.}
\label{hwd}
\end{figure}

To illustrate this difference in more detail, let us consider a copy of the split Cayley hexagon of order two embedded classically in $\mathcal{W}(5,2)$. The 135 planes of the latter space then split into two disjoint, unequally-sized families having 63 and 72 elements.
Every plane of the first family originates, as a point set, from the \emph{perp-set} of a point of the hexagon, 
i.\,e. the set of points collinear with a given point (called the \emph{center}/\emph{nucleus}), the point itself inclusive (and henceforth referred to as a \emph{perp-plane}). 
The planes of the second family form in the hexagon 36 pairs, each pair -- together with corresponding parts of lines of the hexagon -- representing  a copy of the Heawood graph~\cite{heaw} (a \emph{Heawood plane}); 
an illustration of such a pair of Heawood planes is given in Figure~\ref{hwd}, which employs the most common, seven-fold-symmetric rendering of the Heawood graph.

\begin{figure}[t]
\vspace*{-0.cm}
\centerline{\includegraphics[width=10cm,clip]{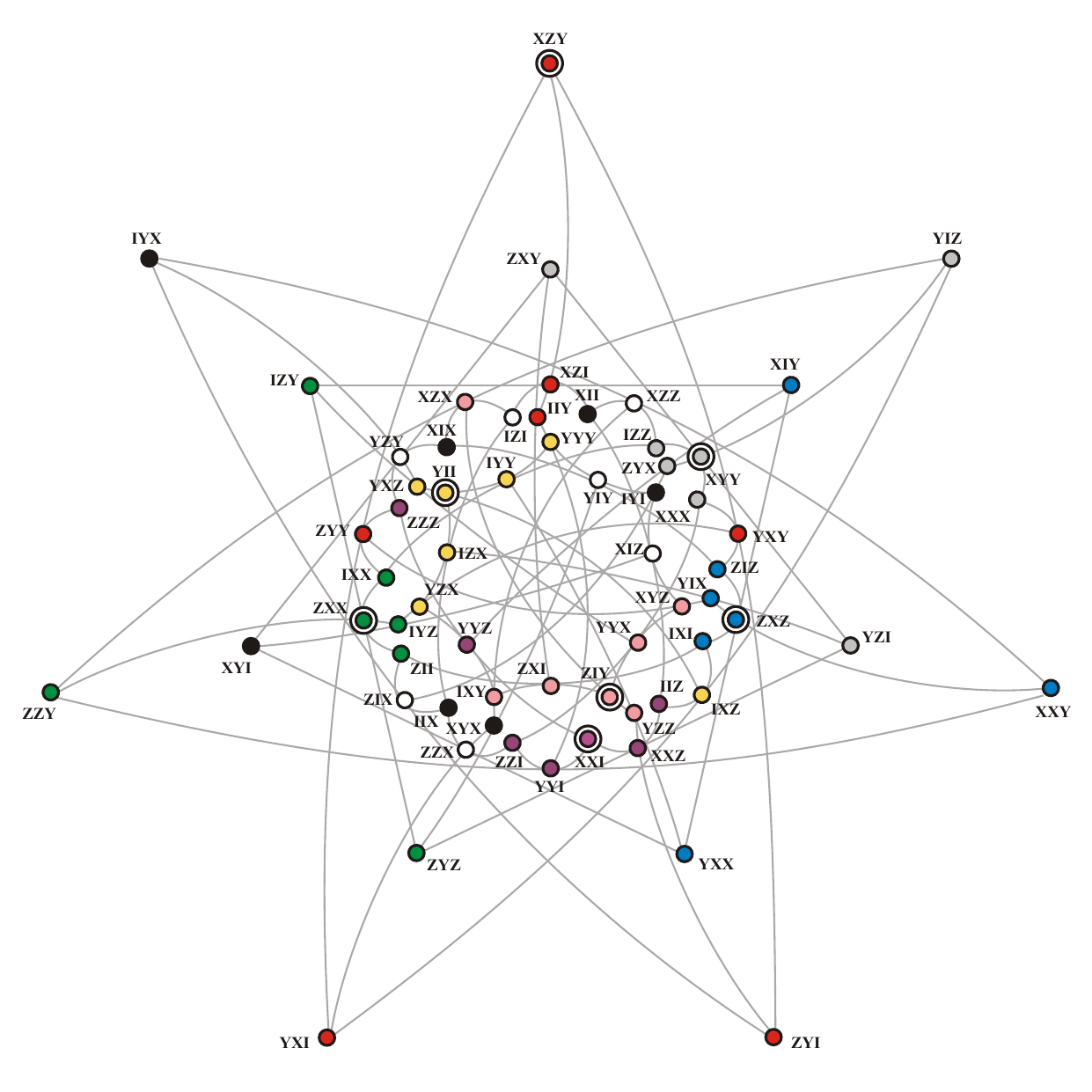}}
\caption{An  illustration of a spread of planes of the first kind. The individual planes (as point sets) are distinguished from each other by different colors, the two planes of Heawood type being represented by black and white (empty) bullets; the nuclei/centers of the perp-sets are encircled. This drawing of the split Cayley hexagon of order two, in a form showing its automorphism of order seven, is a slight modification of those given in~\cites{schr,psm};
 here, a point of the hexagon is represented by a small circle and a line -- each accommodating a triple of pairwise commuting observables whose product is proportional to $I^{\otimes 3}$ -- either by a straight segment (seven of them) or by an arc of variable size and curvature.  Labeling by the three-qubit observables is taken from~\cite{lsv}; see also~\cites{sppl,psh}. 
}
\label{class-1st-kind}
\end{figure}

Similarly, we also find two different kinds of \emph{spreads} of planes, i.\,e. sets of nine pairwise disjoint planes, of $\mathcal{W}(5,\,2)$ with respect to our hexagon. 
A spread of the first kind consists of seven perp-planes and two Heawood planes, the latter coming from the same Heawood graph.  
There are altogether $288$ spreads of this kind. An example, illustrated in a colorful form in Figure~\ref{class-1st-kind}, is furnished by
\begin{eqnarray*}
&& \{\underline{XZY}, ZYY, YXI, YXY, ZYI, XZI, IIY\},\\
&& \{\underline{YII}, IYY, YYY, IZX, YZX, YXZ, IXZ\},\\
&& \{\underline{ZXX}, IXX, ZII, IYZ, ZZY, IZY, ZYZ\},\\
&& \{\underline{ZXZ}, IXI, ZIZ, YIX, XXY, XIY, YXX\},\\
&& \{\underline{ZIY}, ZXI, IXY, YYX, XYZ, YZZ, XZX\},\\
&& \{\underline{XXI}, YYI, ZZI, XXZ, IIZ, YYZ, ZZZ\},\\
&& \{\underline{XYY}, ZYX, YIZ, IZZ, XXX, ZXY, YZI\},\\
&& \{ZZX, ZIX, IZI, XZZ, YIY, XIZ, YZY\},\\
&& \{IYX, XYI, IIX, XYX, XIX, XII, IYI\}; 
\end{eqnarray*}
here, the first seven planes are perp-planes, with the underlined first elements being the nuclei/centers of the corresponding perp-sets.

\begin{figure}[t!]
\centerline{
\includegraphics[width=8.5cm,clip=]{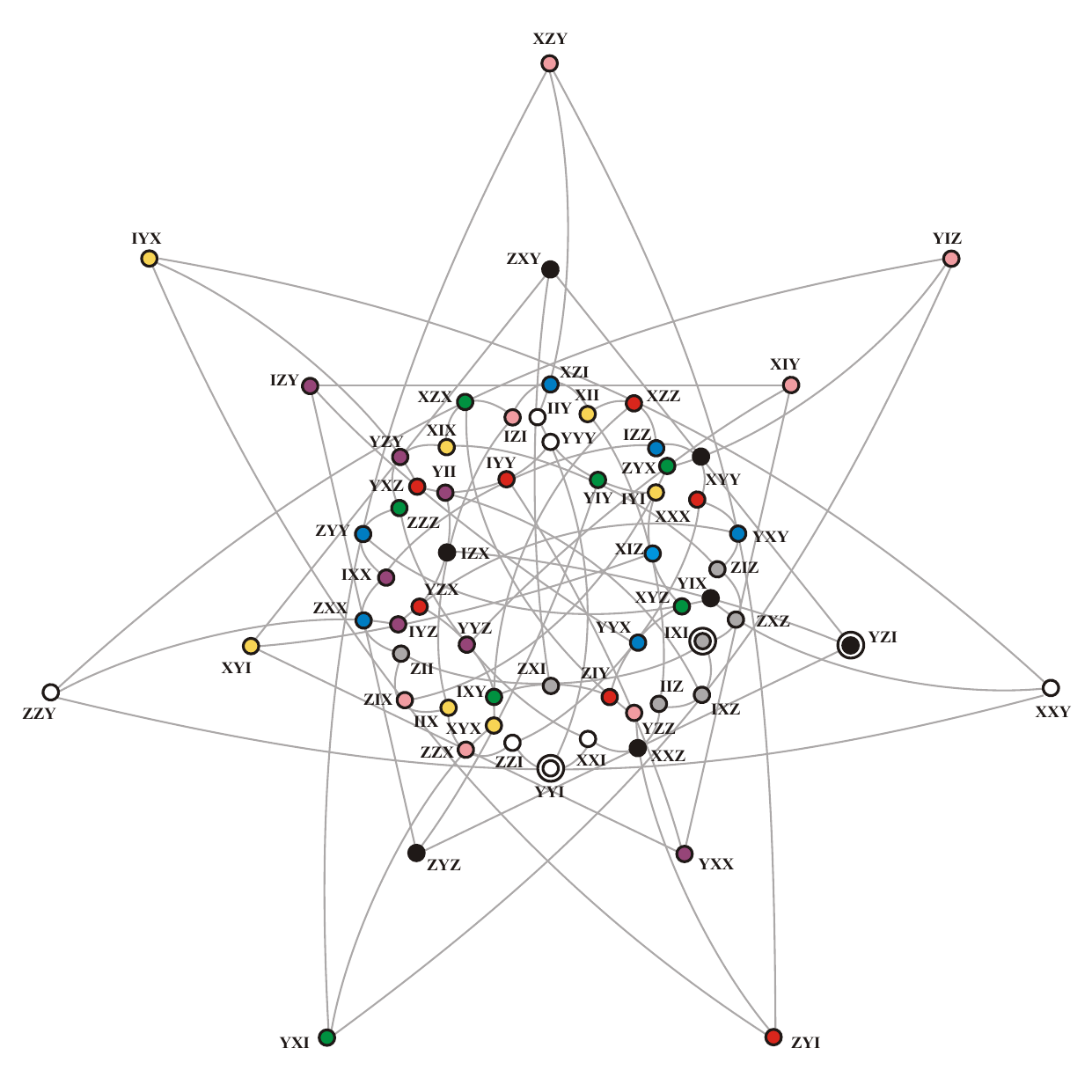}\hspace*{.2cm}\includegraphics[width=8.5cm,clip=]{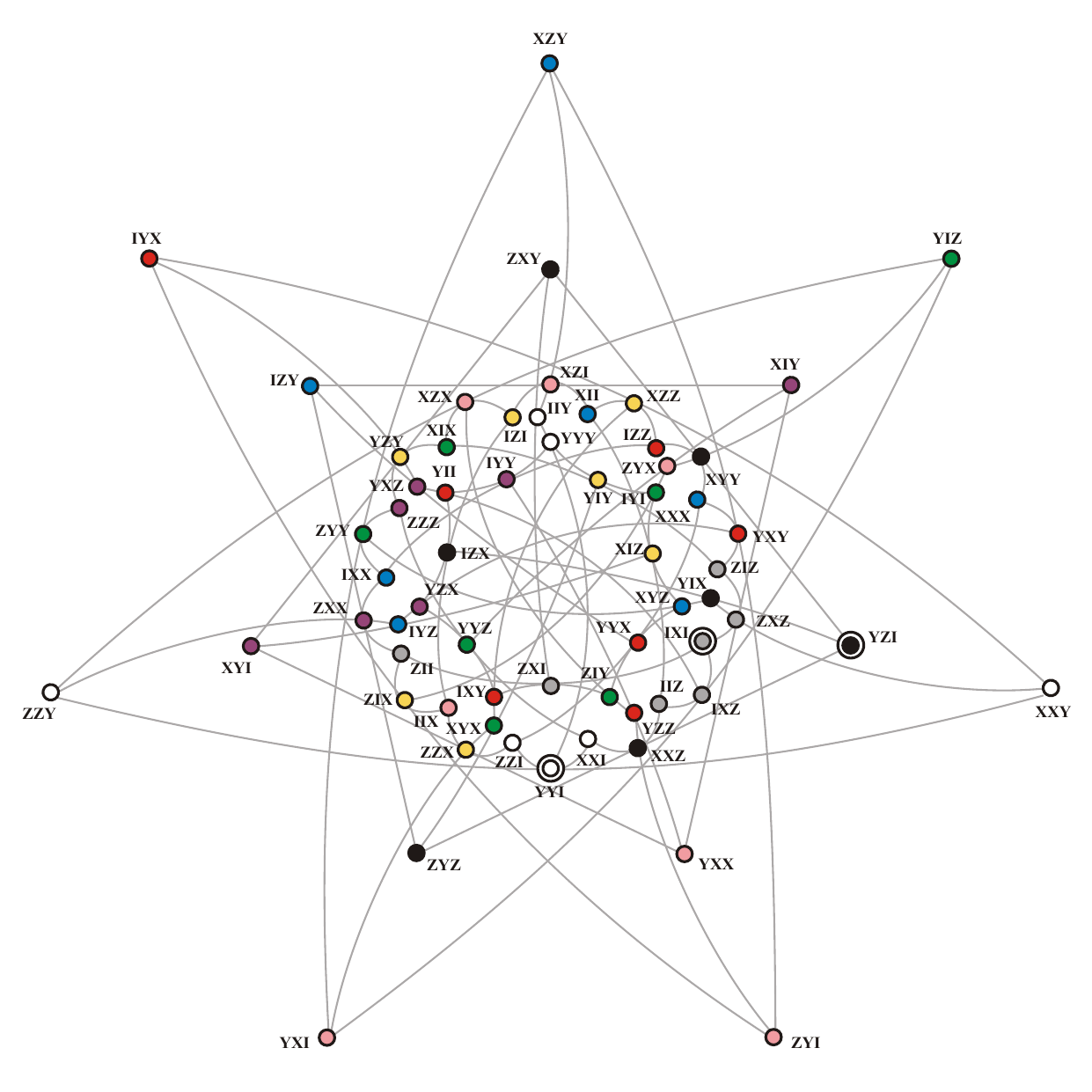}}
\caption{A diagrammatical visualization of the structure of the two spreads of second kind that share three perp-planes, the latter being represented by black, gray and white (empty) bullets; like in the previous figure, the nuclei/centers of the corresponding perp-sets are encircled.}
\label{class-2nd-kind}
\end{figure}

A spread of the second kind features three perp-planes and six Heawood planes, no two of the latter sharing the same Heawood graph. As any three mutually disjoint planes of $\mathcal{W}(5,\,2)$ belong to two distinct spreads, through our three perp-planes goes another spread of the same kind; its six Heawood planes are nothing but the complements of the former six planes in the corresponding six Heawood graphs.  
We have $672$ spreads of this second kind.  Here is a particular pair of spreads of this kind on the same triple of perp-planes, the latter being listed first:
\begin{eqnarray*}
&& \{\underline{YZI}, ZXY, XYY, YIX, IZX, XXZ, ZYZ\}, \\ 
&& \{\underline{IXI}, IXZ, IIZ, ZXZ, ZIZ, ZXI, ZII\}, \\
&& \{\underline{YYI}, ZZI, XXI, ZZY, XXY, YYY, IIY\}, \\
&& \{IYX, XYI, IIX, XYX, XII, XIX, IYI\}, \\
&& \{XZY, YIZ, XIY, YZZ, ZZX, ZIX, IZI\}, \\
&& \{IZY, YZY, YII, IXX, IYZ, YYZ, YXX\}, \\
&& \{XZX, ZZZ, IXY, XYZ, YIY, ZYX, YXI\}, \\
&& \{XZI, ZYY, ZXX, YYX, XIZ, YXY, IZZ\}, \\
&& \{YXZ, YZX, ZYI, ZIY, XXX, XZZ, IYY\}, \\
\end{eqnarray*}
and
\begin{eqnarray*}
&& \{\underline{YZI}, ZXY, XYY, YIX, IZX, XXZ, ZYZ\}, \\ 
&& \{\underline{IXI}, IXZ, IIZ, ZXZ, ZIZ, ZXI, ZII\}, \\
&& \{\underline{YYI}, ZZI, XXI, ZZY, XXY, YYY, IIY\}, \\
&& \{YZY, IZI, XZZ, YIY, XIZ, ZZX, ZIX\}, \\
&& \{XZX, XZI, ZYX, YXX, ZYI, YXI, IIX\}, \\
&& \{XYI, ZXX, YZX, ZZZ, YXZ, IYY, XIY\}, \\
&& \{XIX, ZYY, YYZ, XYX, ZIY, IYI, YIZ\}, \\
&& \{IZY, IXX, IYZ, XYZ, XXX, XII, XZY\}, \\
&& \{IYX, YII, IXY, YZZ, YYX, YXY, IZZ\}. \\
\end{eqnarray*}
This particular pair of spreads is also illustrated in Figure~\ref{class-2nd-kind}; here the union of any two Heawood planes represented by the same color forms a copy of the Heawood graph.

In a skew-embedded hexagon, however, {\it neither} of the above-described two patterns can be found; this is mainly due to an {\it in}sufficient number of planar points, but also due to the {\it way} how these points are arranged with respect to each other.

\subsection{Skew-embedded hexagon, linear doilies and contextuality}\label{subsec:linear}
\begin{figure}[pth!]
\vspace*{-2cm}
\centerline{\includegraphics[width=15cm,clip=]{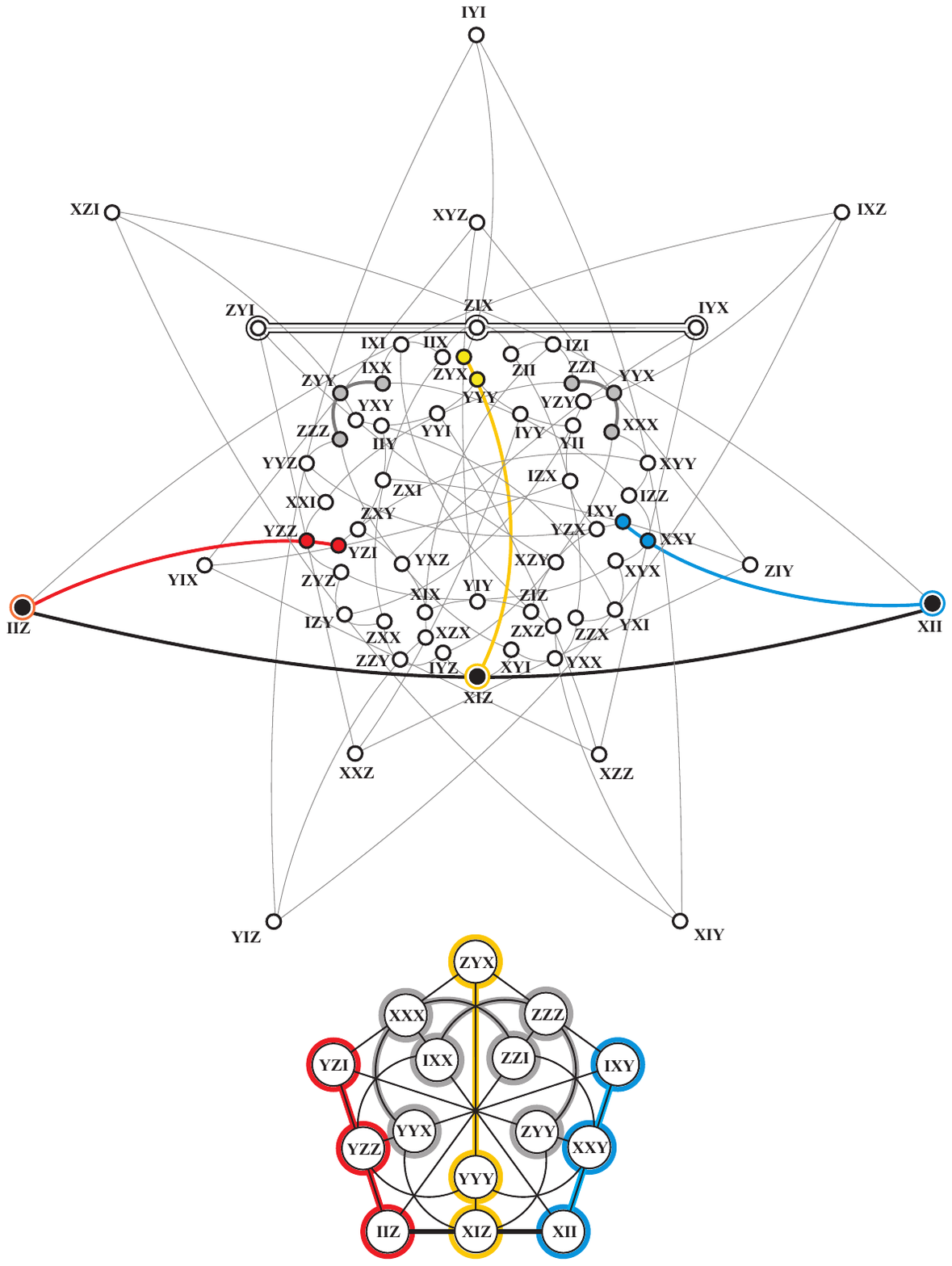}}
\caption{An illustration of the procedure that shows that to each `non-planar' line (black) of a skew-embedded
split Cayley hexagon of order two (top) one can associate a unique linear doily (bottom) that shares with it the maximum possible number of lines. The horizontal highlighted line of the hexagon is its axis; the meaning of the remaining colored lines is explained in the text.}
\label{skew-to-doily}
\end{figure}

The fact that a copy of the split Cayley hexagon of order two embedded skewly into $\mathcal{W}(5,2)$ features non-planar points
has a number of interesting consequences. We will briefly address one of them. Given a skew-embedded
hexagon, like the one depicted in Figure~\ref{skew-to-doily}, let us pick up in it a line that consists solely of non-planar points, for example the line $IIZ-XIZ-XII$ (drawn black in Figure~\ref{skew-to-doily}).  Through each point of this line there pass one line that does not belong
to the plane of $\mathcal{W}(5,2)$ defined by the other two  concurrent lines at this point; these are the lines $IIZ-YZZ-YZI$ (via point
$IIZ$ -- red), $XIZ-YYY-ZYX$ (via $XIZ$ -- yellow) and $XII-XXY-IXY$ (via $XII$ -- blue).
Clearly, these three lines are pairwise disjoint as otherwise our generalized hexagon would contain triangles, which is impossible as its smallest ordinary polygons are hexagons. 
These three lines define a unique {\it linear} doily (i.\,e. a doily located in a certain PG$(3,2)$ of the ambient PG$(5,2)$), namely the one depicted at the bottom of Figure~\ref{skew-to-doily}. However, this doily shares with our hexagon two more lines (shown in gray) that are disjoint from each other and also from any of the three lines. 
It is obvious that such a set of six lines is the {\it maximum}
set of lines a doily and a skew-embedded hexagon can share; indeed, assuming that our doily shares an additional line with the hexagon that is skew to the black line (i.\,e. not incident with it) would mean that the latter would contain quadrangles, which contradicts its definition.

The above-described relation can also help us understand why a classically-embedded hexagon and a doily share just three lines
belonging to a grid of the doily. For if we disregard in Figure~\ref{skew-to-doily}  the three colored lines that occur only
in skew-embedded hexagons, the remaining three lines (i.\,e. the black line and the two gray lines) indeed belong to a particular grid of the doily in question!

\begin{figure}[ht]
\vspace*{-0.cm}
\centerline{\includegraphics[width=12cm,clip=]{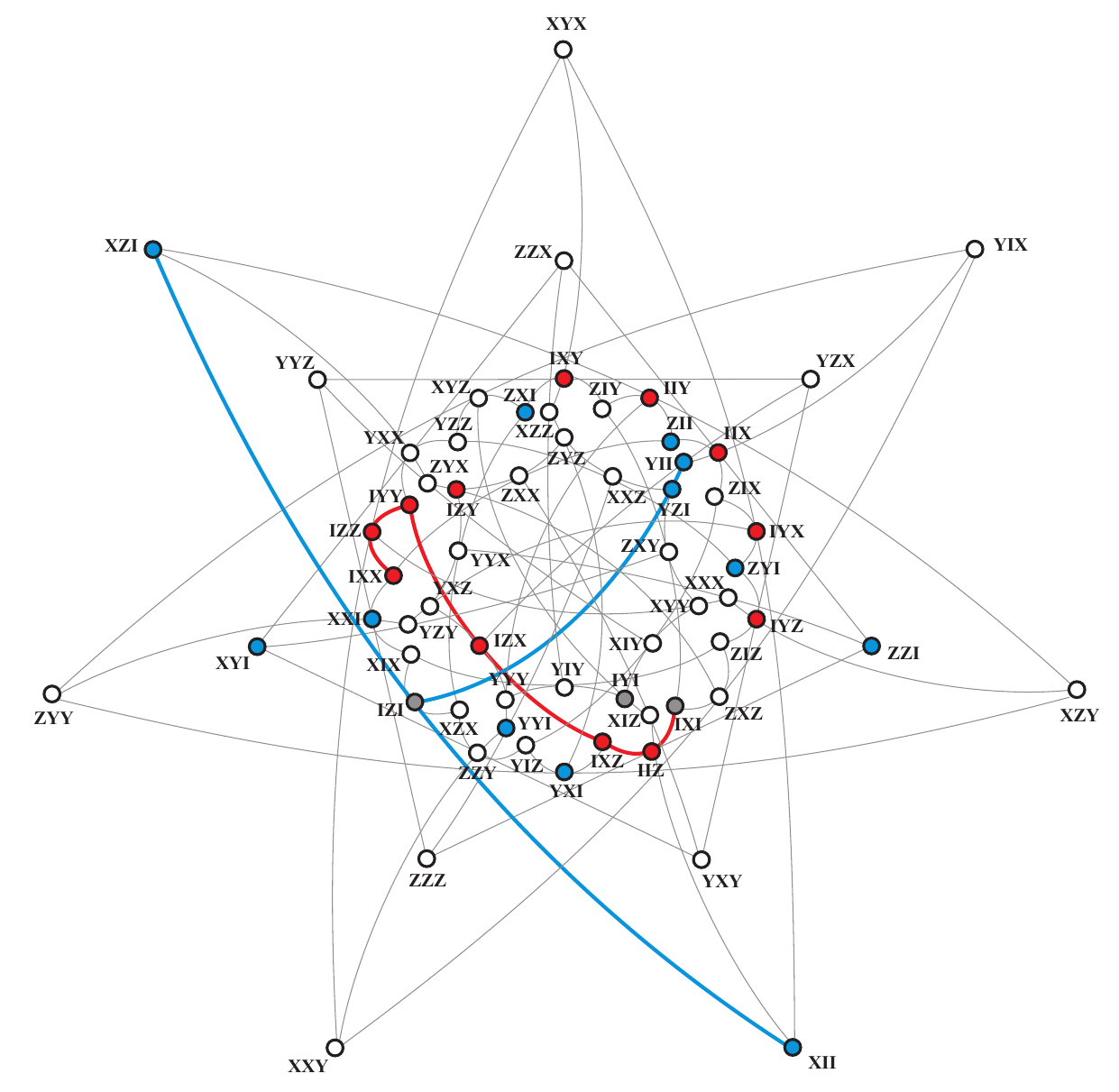}}
\caption{A different copy of skew-embedded hexagon (whose axis is the line $YYZ-IXY-YZX$) employed to illustrate additional doily-hexagon
intersection patterns. The points and (hexagon-shared) lines of the two doilies are highlighted in different colors, 
the three points that belong to both the doilies being colored gray.}
\label{skew-A}
\end{figure}

By a computer search we have found out that there are only three more patterns of lines a linear doily can share with a skew-embedded hexagon. One of them is an already mentioned set of three mutually disjoint lines that belong to some grid of the doily. There exists another three-line pattern, namely that comprising two disjoint lines having a common transversal; slightly rephrased, this pattern entails any three lines forming sides of a quadrangle in the doily. The remaining type features two intersecting lines. These can readily be illustrated employing the copy of skew-embedded hexagon depicted in Figure~\ref{skew-A}. To this end, let us consider three particular linear three-qubit doilies, namely the doilies whose all 15 observables feature $I$ on the first qubit (the `left' doily), on the second qubit (the `middle' doily) and on the third qubit (the `right' doily). From Figure~\ref{skew-A} one can easily discern that the `left' doily (red) shares with the hexagon the following three lines $IXX-IZZ-IYY$, $IYY-IZX-IXZ$ and
$IXZ-IIZ-IXI$, the middle one being indeed incident with either of the remaining two that are disjoint. On the other hand, the `right' doily (blue) has two concurrent lines in common with the hexagon: $XZI-IZI-XII$ and $IZI-YZI-YII$. For the sake of completeness, we also mention  that the  `middle' doily (not shown) shares with the hexagon the maximal pattern, composed of the five  lines $ZIY-IIY-ZII$, $YII-IIX-YIX$, $ZIX-XIY-YIZ$, $XIX-YIY-ZIZ$ and $XIZ-IIZ-XII$ forming a spread, and the line $ZII-IIX-ZIX$.

The above-described intersection patterns of linear doilies with skew-embedded hexa\-gons are of crucial importance for a better understanding of the fact why the complements of the latter are contextual. This basically boils down to the fact that a less symmetric, skew-embedded hexagon just fits into $\mathcal{W}(5,2)$ in such a manner that it provides, in contrast to its much more symmetric, classically-embedded sibling, enough `space' for its complement to contain grids (aka Mermin-Peres magic squares). To see this explicitly, 
let us consider a doily having with a skew-embedded hexagon two concurrent lines in common. It is easy to see
that out of the ten grids in this doily, there are four of them that do not contain any of the two shared lines. That is, these
four grids are {\it fully} contained in the complement of the hexagon. A similar situation also occurs in the above-described  
three-line pattern; in this case the corresponding doily has just two grids devoid any of the three lines and, so, lying fully in the hexagon's complement. As grids are the simplest, and so fundamental, quantum contextual structures in multi-qubit symplectic polar spaces, their occurrence in the complements of {\it skew}-embedded hexagons lends itself as one of the most natural justifications why these complements are contextual.  

\section{The contextuality of \texorpdfstring{$\mathcal{W}(5,2)$}{W52} and the classical embedding of \texorpdfstring{$\mathcal{H}$}{H}}\label{sec:W52}
Having a better understanding of the difference between the two symplectic embeddings
of $\mathcal{H}$ and being endowed with a fresh insight on why the complement of a skew-embedded $\mathcal{H}$ is contextual (by containing three-qubit grids), we can now address our main objective: the relevance of hexagon's classical embeddings for
the issue of three-qubit contextuality.

Recently, the authors of~\cite{muller2023new} made the key (computer-based) discovery  that the degree of contextuality of the configuration comprising all 315 line-contexts of $\mathcal{W}(5,2)$ is equal to $63$, and not $90$ (= the number of negative lines) as previously proposed by~\cite{Cab10}. In addition, and more importantly, they found out that the 63 unsatisfiable contexts of this configuration are in bijection with 63 lines of a copy of $\mathcal{H}$ that is embedded {\it classically} into $\mathcal{W}(5,2)$. These facts came as a big surprise to us and prompted us to   
have a closer look at what is going on here. We will provide first some algebraic-geometrical arguments for the occurrence of the number $63$. Then, employing a line-layered decomposition of the hexagon, we will demonstrate on several examples how the contextuality properties of a three-qubit configuration having three-element contexts can simply be read off from its generic intersection with a classically-embedded copy of $\mathcal{H}$.

\subsection{The degree of contextuality of \texorpdfstring{$\mathcal{W}(5,2)$} is sixty-three}

In~\cite{muller2023new} the proof that the degree of contextuality of $\mathcal{W}(5,2)$ is $63$ was obtained by computer. After  translating the problem of finding an NCHV model into the resolution of a linear system over $\mathbb{F}_2$, the authors took advantage of a SAT solver to provide an explicit model where all but $63$ of the $315$ constraints imposed by the observable-labelled lines of $\mathcal{W}(5,2)$ are satisfied. This existence of an explicit model proves that the degree of contextuality $d$ of $\mathcal{W}(5,2)$ satisfies $d\leq 63$. In~\cite{muller2023new} the inequality $d\geq 63$ was deduced from the fact that the SAT solver was unable to find an explicit NCHV model  with at most $62$ constraints, showing $d\geq 63$.   It turns out that this second inequality can be obtained from a tiling of the lines of $\mathcal{W}(5,2)$ by doilies. Recall~\cite{sbhg21} that $\mathcal{W}(5,2)$ features two kind of doilies, referred to as linear and quadratic. One can define (see Section 2 of~\cite{sbhg21}) a quadratic doily as the intersection of a hyperbolic quadric and an elliptic one in $\mathcal{W}(5,2)$. 
From eqs.~\eqref{hqinwn} and~\eqref{eqinwn} in Section~\ref{sec:degree}  for $n = 3$ 
it follows that there are  $36$ quadrics of the former and $28$ of the latter type, which yields $36\times 28=\numprint{1008}$ quadratic doilies in total. The symplectic group, $Sp(6,2)$, acts transitively not only on the lines of $\mathcal{W}(5,2)$, but also on pairs of quadrics. 
The geometry of these specific pairs of quadrics, whose intersection defines a quadratic doily, has been investigated in full detail in~\cite{levay2017magic}. Each line of $\mathcal{W}(5,2)$ is shared by  $48$ different doilies and, by transitivity of $Sp(6,2)$, the $\numprint{1008}$ quadratic doilies cover all the $315$ lines/contexts of $\mathcal{W}(5,2)$. Now recall (see Section~\ref{sec:degree}) that each doily features three constraints that cannot be satisfied. So the restriction to a doily of an NCHV model of $\mathcal{W}(5,2)$ should induce at least $3$ constraints on each doily of $\mathcal{W}(5,2)$. This implies that the degree of contextuality $d$ of $\mathcal{W}(5,2)$ should satisfy the following inequality:
\begin{equation}
d\geq\dfrac{\numprint{1008}\times 3}{48}=63.
\end{equation}
Similar group-theoretic arguments can be used if we consider the tiling of lines by other contextual subgeometries of $\mathcal{W}(5,2)$. For instance, the $336$ linear doilies also cover all the $315$ lines of $\mathcal{W}(5,2)$, and as each line of $\mathcal{W}(5,2)$ sits in $16$ linear doilies, we arrive at the same result: $d\geq\dfrac{336\times 3}{16}=63$.

The above-given group-combinatorial explanation of the number 63 can further be substantiated geometrically in the following sense: given a {\it classically}-embedded $\mathcal{H}$ one can find a set of 21 doilies whose 63 shared lines (three per doily) {\it partition} the set of lines of this $\mathcal{H}$. Next, the fact that a classical
$\mathcal{H}$ encodes quantitative information about the contextuality of $\mathcal{W}(5,2)$ {\it  as a whole} invokes the thought that this encoding should also manifest
on {\it any} contextual subgeometry of $\mathcal{W}(5,2)$ whose contexts are lines.
And this is indeed the case. However, to see it explicitly, we still need to delve a bit more into the geometric structure of the hexagon.

\subsection{Layering of the split Cayley hexagon of order two}
\begin{figure}[pth!]
\centerline{\includegraphics[width=15truecm,clip=]{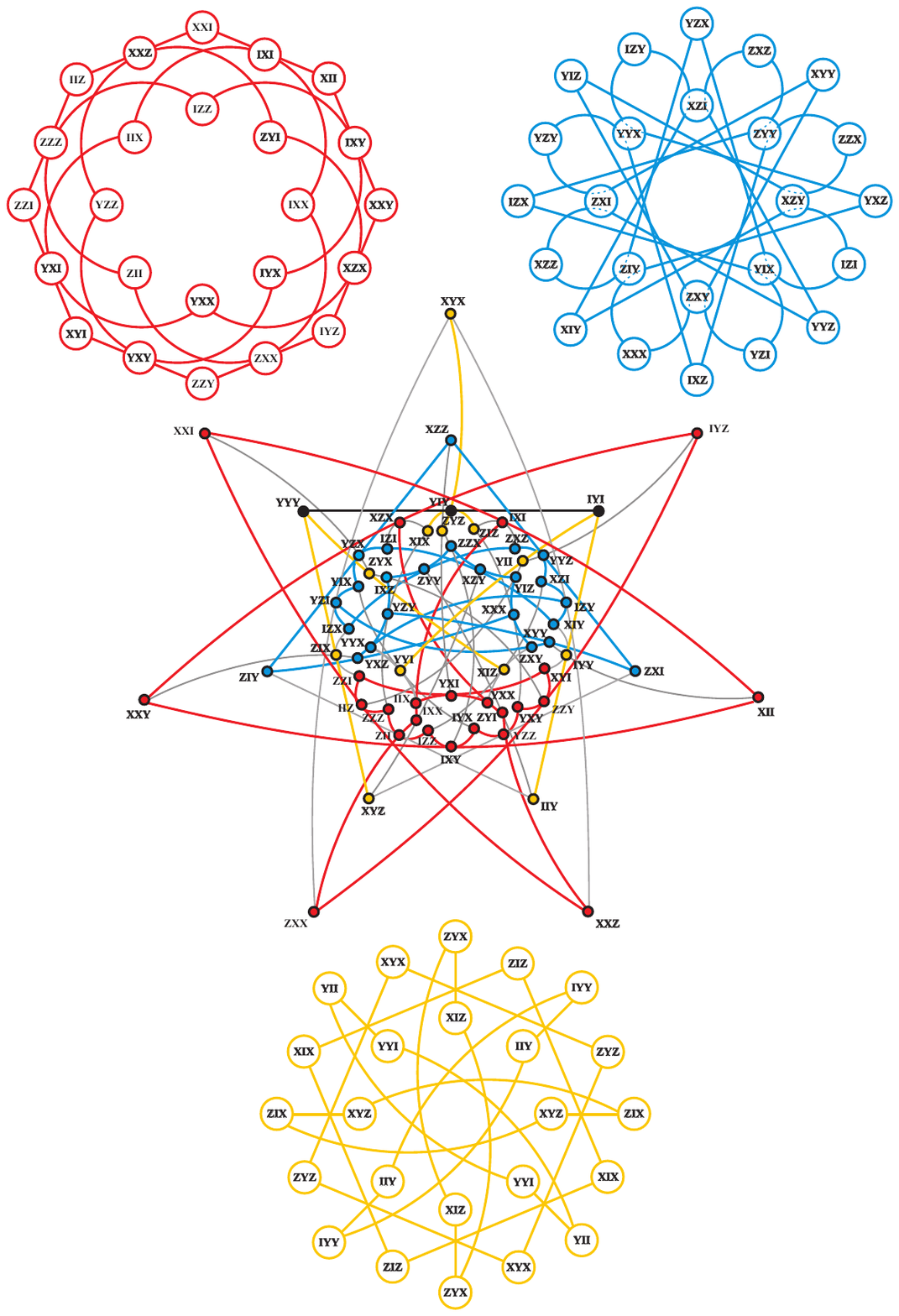}}
\vspace*{.2cm}
\caption{An illustration of the line-distribution (i.\,e., the distribution of contexts) within the hexagon.}
\label{lay-hex}
\end{figure}

Let us pick up a skew-embedded copy of the hexagon, e.g. the one shown in Figure~\ref{lay-hex}, and
highlight its {\it axis} by black color. Given this line, the remaining 62 lines split into
three distinct sets. The first set comprises six lines (yellow ones), each being incident
with the axis. The second set features 24 lines (gray), each being incident with some yellow
one. The last set consists of the remaining 32 lines; these lines can be divided into
two {\it disjoint}, isomorphic sets of  16 elements each -- the two sets being distinguished by blue and
red colors.

To see finer traits of the structure of the hexagon, one can redraw the three main 
`layers' or `domains' of the hexagon,
namely red (top left), blue (top right) and yellow (bottom) in the most symmetric way.
Note that both the red and blue domains can each be viewed as a pair of circumscribed
octagons, either of them being isomorphic to the same $(24_2, 16_3)$-configuration. In the
illustration of the yellow domain each observable is represented by two different
(opposite) points and the corresponding \emph{affine} part of each yellow line, 
i.\,e. that two-point part of the line that is left after the removal of the point lying on the axis, has four distinct images forming
a quadrangle. Our option for such a  rendering of the yellow layer is simple: if one stacks
all the three figures above each other then the corresponding three observables at a given
position define a particular gray line of the hexagon; for example, the three topmost
observables form the gray line $XXI-ZYX-YZX$.

\subsection{A simple recipe how to get from a skew-embedded hexagon a classically-embedded one, and vice versa}\label{sec:1.2}
\begin{figure}[t]
\vspace*{-2.0cm}
\centerline{\includegraphics[width=11truecm,clip=]{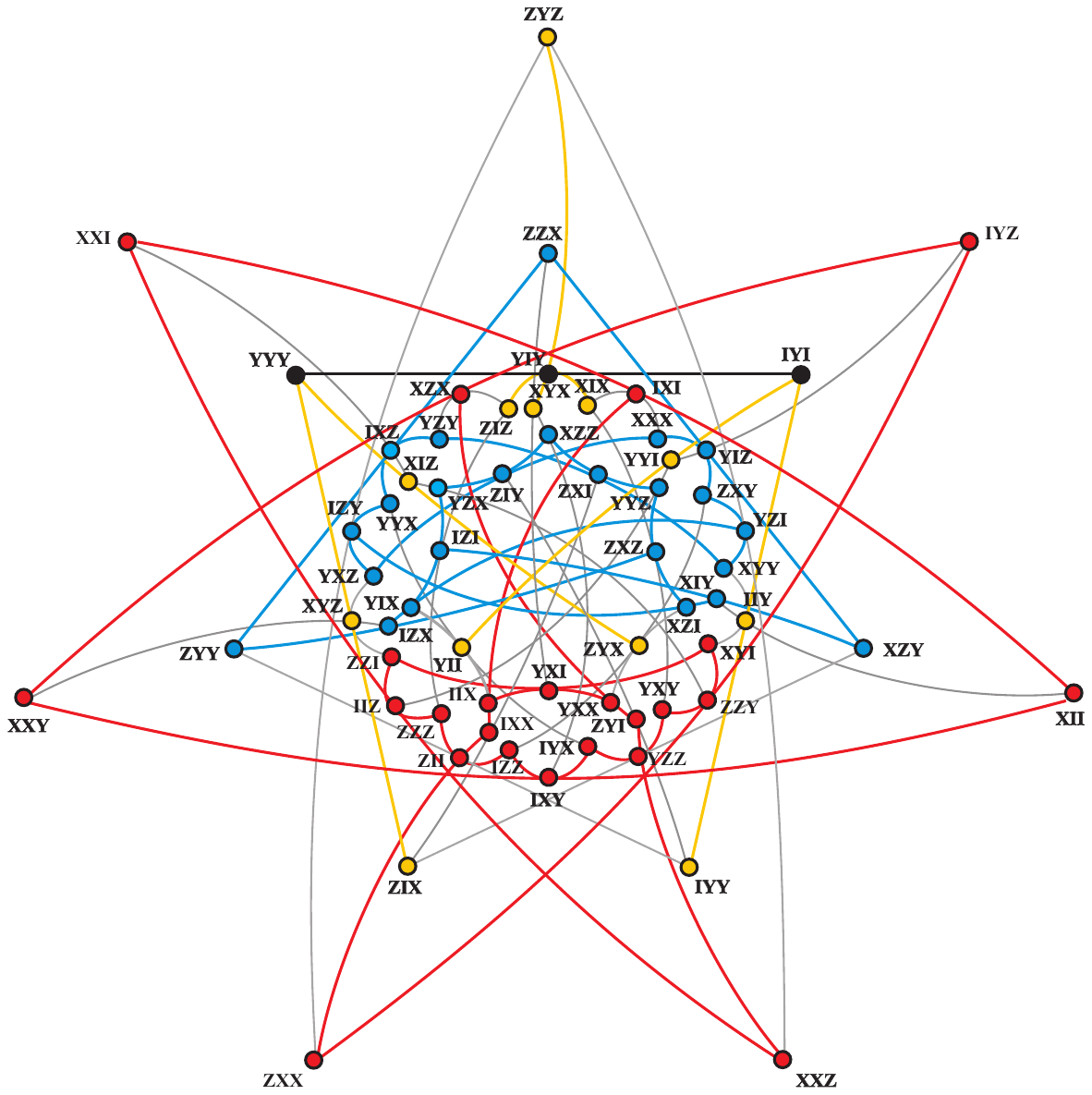}}
\vspace*{-3.0cm}
\caption{The classical hexagon we get from the skew-embedded one by applying the
procedure described in the text.}
\label{clemb-hex}
\end{figure}

The above-described layered structure of the hexagon turns out to be very relevant to better
understand the relation between the two types of embeddings of the split Cayley hexagon of order two. 
In fact, there exists a rather simple recipe that `transforms' a skew hexagon into a classical one.
Let us start with the skew hexagon shown in Figure~\ref{lay-hex}.
Keep the three observables of the axis intact. On each yellow line, {\it swap} the two
remaining observables. On each red (blue) line leave each observable intact and on each
blue (red) line replace its observable by that which is the product of a swapped yellow
observable and a fixed red (blue) observable on the corresponding gray line. What we get is a
copy of a classically-embedded hexagon, depicted in Figure~\ref{clemb-hex}, that shares 39 lines with the original
skew hexagon: the black line, the 6 yellow lines, the 16 blue lines and the 16 red lines.

This construction {\it only} works if the black (reference) line is the axis of the skew hexagon.
Obviously, we can reverse the process and start with a classically-embedded hexagon to get a skew one.
In this case {\it any} of its 63 lines can be taken as the reference black line, so we get
63 different skew copies from a given classical one. As there are 120 different classical copies of the hexagon in
$\mathcal{W}(5,2)$, the above property also implies that there are as many as $120 \times 63 = 7560$ skew-embedded hexagons 
living in $\mathcal{W}(5,2)$ -- confirming the result of~\cite{holweck2022three} based on an exhaustive computer search.

\subsection{Classically-embedded hexagons encode three-qubit contextuality}
\label{sec:class3qubit}

As already stressed, the  fact that any set of 63 unsatisfiable
constraints associated with all the 315 lines/contexts of the three-qubit symplectic polar space forms a $63_3$-configuration
isomorphic to a copy of the split Cayley hexagon of order two embedded {\it classically} into the space in question seems to be just part
of a bigger story, as something similar is taking place for the lines located on both
elliptic and hyperbolic quadrics, as well as on doilies of
$\mathcal{W}(5,2).$
 
\begin{figure}[t]
\vspace*{-.0cm}
\centerline{\includegraphics[width=11truecm,clip=]{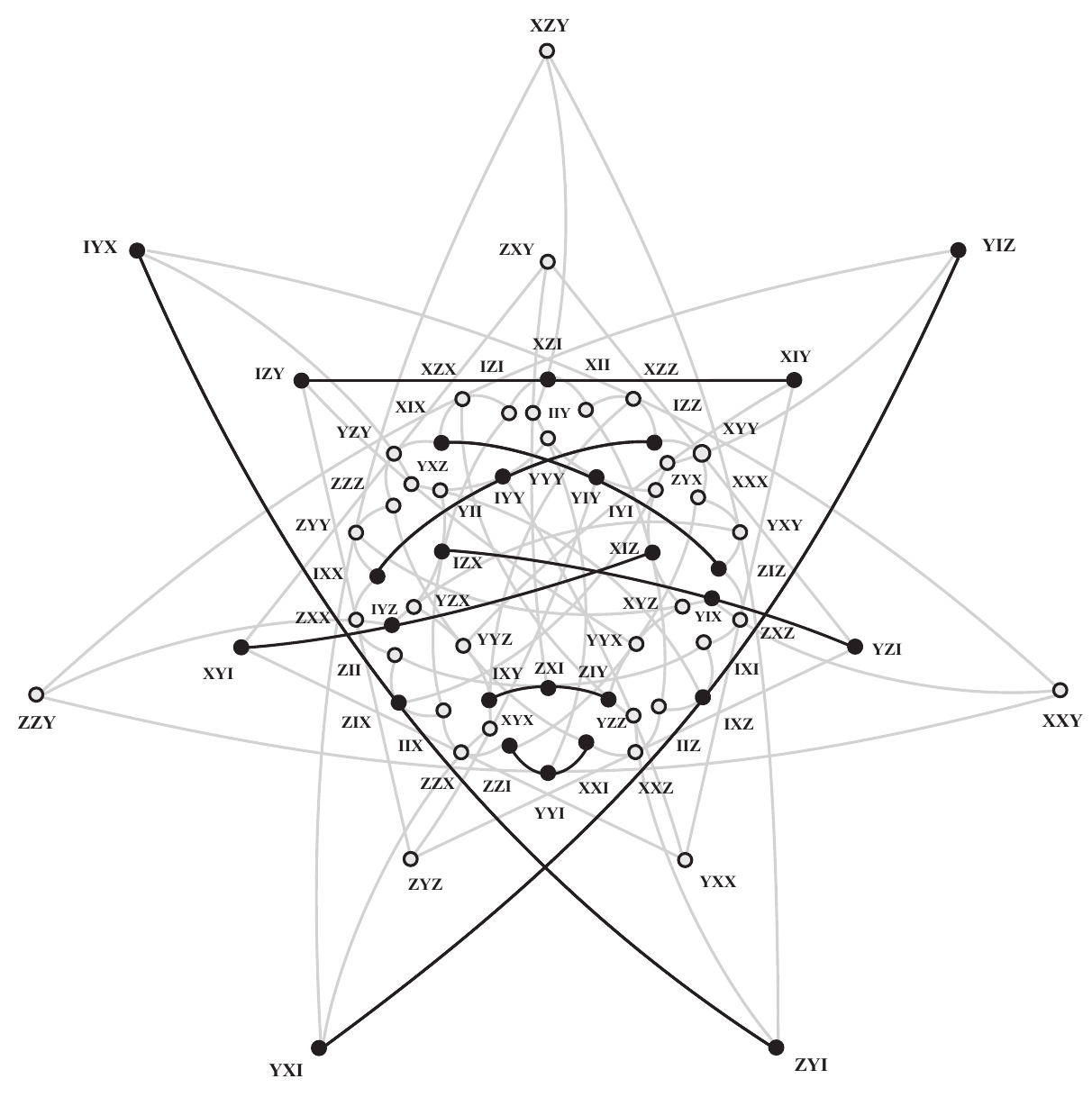}}
\caption{The generic intersection (bold-faced points and lines) of a classical hexagon with the elliptic quadric whose index is $YYY$: nine mutually disjoint lines that are at maximum distance from each other. Note that these nine lines cover all the 27 points of the quadric.}
\label{hex-ell}
\end{figure}

\begin{figure}[t]
\vspace*{-0.cm}
\centerline{\includegraphics[width=11truecm,clip=]{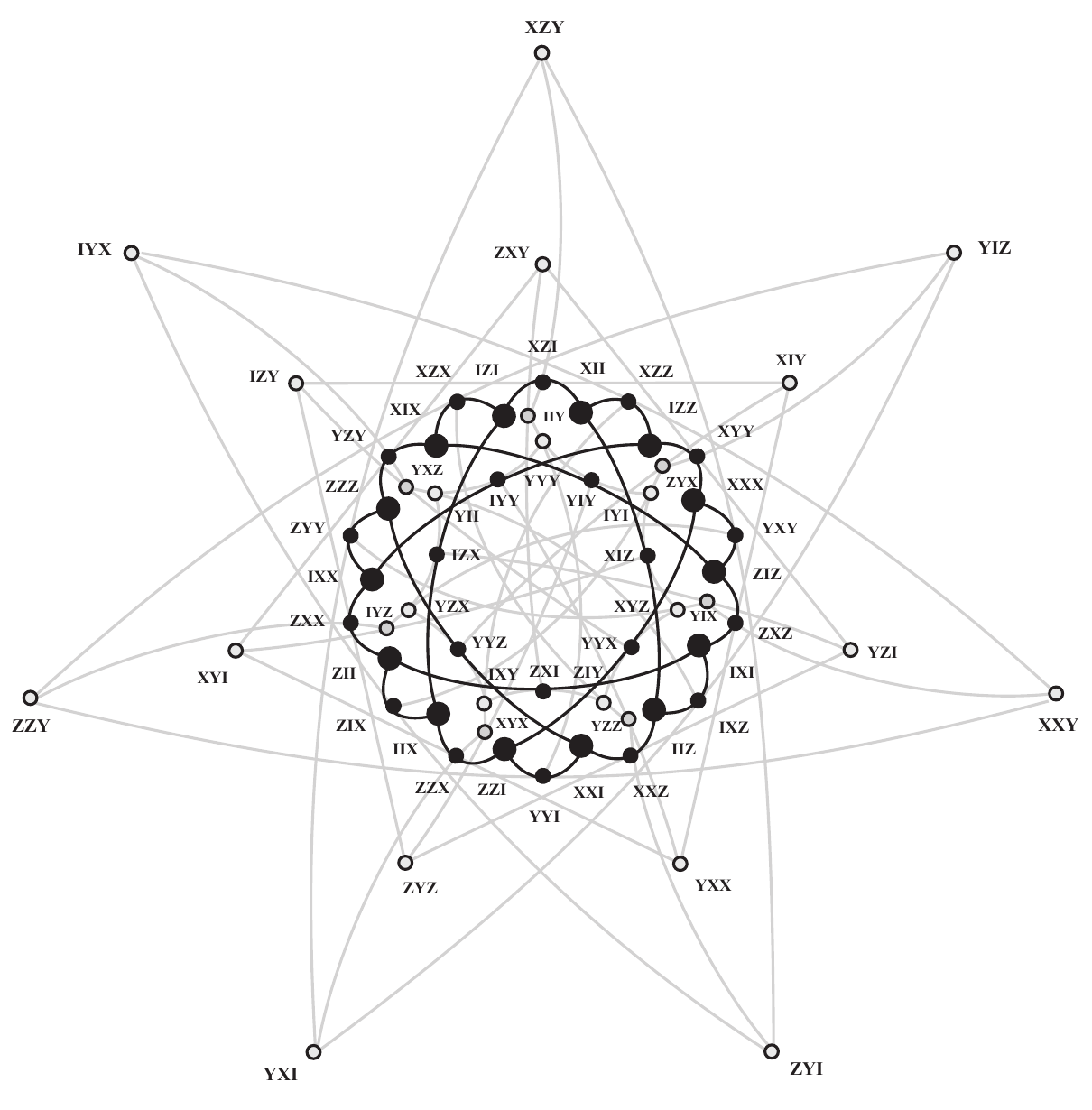}}
\caption{The generic intersection (bold-faced points and lines) of a classical hexagon with the hyperbolic quadric whose index is $III$: 21 lines forming edges of the Heawood graph (compare also with Figure~\ref{hwd}). As in the previous case, the 21 lines comprise all the 35 points of the quadric.}
\label{hex-hyp}
\end{figure}

Let start with elliptic quadrics. By a computer search we have found that each elliptic quadric features
9 pairwise disjoint unsatisfiable lines/contexts forming a spread. On the other hand, each classically-embedded 
hexagon shares with each elliptic quadric such a set of 9 lines. An example is illustrated in
Figure~\ref{hex-ell}.
Next, each classically-embedded hexagon shares with each hyperbolic quadric exactly 21 lines, forming a pattern 
isomorphic to the one shown in Figure~\ref{hex-hyp}. 
By comparing this figure with Figure~\ref{hwd} one readily recognizes
this pattern as the Heawood graph, also known as the point-line incidence graph of the Fano plane,  whose vertices are represented by bigger bullets, each of its edges being 
supplied with one more point/observable to
represent a full line of the three-qubit $\mathcal{W}(5,2)$.  
Using a computer-aided search based on a SAT solver we have verified that each three-qubit hyperbolic quadric features 21 unsatisfiable lines whose arrangement is isomorphic to that depicted in Figure~\ref{hex-hyp}.
A similar analysis carried out for doilies living in $\mathcal{W}(5,2)$ showed the same result. The three unsatisfiable constraints a doily features
are those three lines that the doily shares with a particular copy of the classical hexagon. 
An apt example of this property is furnished by the quadratic doily that is shared by the elliptic quadric of index $YYY$ and the hyperbolic quadric
of index $III$. By comparing Figure~\ref{hex-ell} and Figure~\ref{hex-hyp} one can see that there are indeed just three boldfaced lines that occur in both the figures, namely the lines $IXX-IYY-IZZ$, $XIX-YIY-ZIZ$ and $ZZI-YYI-XXI$.

The message from the above-given observations is more than obvious: the {\it degree of contextuality} of a
contextual quantum geometry of the three-qubit $\mathcal{W}(5,2)$ whose contexts are i) all the 315 lines of the whole space $\mathcal{W}(5,2)$, ii) all the 45 lines of
an elliptic quadric, iii) all the 105 lines of a hyperbolic quadric or  iv) all the 15 lines of a doily is equal to the {\it number
of lines} each of these  geometries shares with a classically-embedded hexagon, with the understanding that each shared line corresponds to an unsatisfiable context. A natural generalization of these findings is
that this should hold for {\it any} contextual sub-geometry of $\mathcal{W}(5,2)$ whose contexts are {\it lines}.  \ul{In other words, classically-embedded copies of $\mathcal{H}$, although being {\it non}-contextual by themselves, are found to rule three-qubit contextuality with three-element contexts by being a reliable tool not only to  check whether a particular subgeometry of $\mathcal{W}(5,2)$  is contextual, but also, and still mysteriously, to `extract' from this subgeometry exactly that part that makes it non-contextual!}

As a particular stance of this conjecture, let us consider the complement of a skew-embedded hexagon, which is a contextual $(63_{12}, 252_3)$-configuration~\cite{holweck2022three}. Analyzing complements of several different copies of skew-embedded hexagons we always arrived at the same result: each complement
featured 24 unsatisfiable constraints that were exactly those 24 lines in which the given skew-embedded hexagon differs from the
classical one derived from it using the procedure described in Section~\ref{sec:1.2}; thus, for example, if we consider the complement of the skew hexagon shown in Figure~\ref{lay-hex} then its 24 unsatisfiable lines are exactly the 24 gray lines of its derived classical sibling portrayed in Figure~\ref{clemb-hex}!

\section{Some experimental verification using the IBM Quantum Experience}\label{sec:ibm}

The advent of accessible NISQ computers has allowed researchers to test whether contextual geometries exhibit the predicted quantum behaviour in reality, see for example~\cites{laghaout2022demonstration,holweck2021testing, Kelleher_2_qubit_games,tran22,brav23,alla16,xu_experimental_2022}. 
Given the accelerated development of such technologies, and their widening availability for research purposes, it is easier than ever to experimentally test out the predictions of NCHV models mentioned in the introduction. First, one needs a quantitative test on our geometry that can rule out NCHV models in favour of those predicted by quantum mechanics (QM). In this section we examine such a test. 
\subsection{Cabello inequality}
In 2010 Cabello~\cite{Cab10} introduced an inequality dependent upon the number of satisfiable constraints in a contextual geometry. The quantity $\chi$ is defined as the sum of expectation values of all constraints, with negative ones picking up a sign change, 
\begin{align}
     \chi &\equiv \sum_{i}\langle C_{i}\rangle - \sum_{j} \langle C'_{j} 
 \rangle \\
     \chi &\leq \begin{cases}
         N, \quad \qquad \, \text{QM} \\
         N - 2d, \quad \text{HV} \label{eq:cabello_ineq}
     \end{cases}
 \end{align}
where the $C_{i}$ are the positive contexts in the geometry, $C'_{j}$ the negative ones, $\langle \cdot \rangle$ their expectation values, $N$ the total number of contexts, and $d$ the degree of contextuality. In the QM regime, all constraints are satisfied with expectation value $+1$ for positive contexts and $-1$ for negative ones, and so $\chi$ provides the number of constraints in total. For NCHV models, some constraints will not be satisfied and induce an additive factor of $-2$ into this expression. 

The inequalities provide an upper bound on the measured value of $\chi$ based on whether all constraints are satisfiable (QM) or there is an NCHV model constraining some measurement outcomes (HV).

\subsection{Measuring contextuality on an NISQ computer}
One can experimentally test the contextuality (that is, the value of $\chi$ in~\eqref{eq:cabello_ineq}) of a given labelled geometry via the IBM Quantum Experience~\cite{ibmq}. The methodology is straightforward: for a given context in a geometry labelled by $3$-qubit operators, measure the values of the operators at each point of the context and combine to get the measured parity of the line. Repeat for all lines in the geometry, and sum together weighted by their signed parities. If the final result is greater than the upper bound $N - 2d$ given in~\eqref{eq:cabello_ineq}, then we have demonstrated the contextuality of the geometry.

There is but one issue to address first before implementing this in a circuit. For a context containing three operators $\mathcal{O}_{1},\mathcal{O}_{2},\mathcal{O}_{3}$, when measuring the state of operator $\mathcal{O}_{1}$ one has destroyed the quantum state, and measurements of $\mathcal{O}_{2}, \mathcal{O}_{3}$ will no longer be descriptive of the context as a whole. To circumvent this, we introduce additional ``delegation" qubits into the circuit. The purpose of these qubits is to record the ``state" of the context under each operator, without destroying it via measurements on the original register.

\begin{figure}[ht]
\centering
    \includegraphics[width=0.5\textwidth]{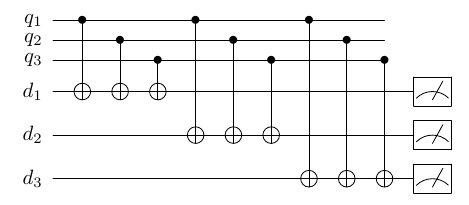}
    \caption{Quantum circuit demonstrating triple measurement on a context via delegations. Qubits $q_{1}, q_{2}, q_{3}$ encode the three-qubit state of the context. State is passed down to delegation qubits $d_{1}, d_{2}, d_{3}$ sequentially via CNOT gates. Basis change gates are suppressed.}
    \label{fig:circuit}
\end{figure}

Firstly, we initialise three qubits labelled $q_{1}, q_{2}, q_{3}$ encoding the state of the three qubits the context will act on. Depending on the operator $\mathcal{O}_{1}$ of the context, gates are applied to these qubits to change their basis from the computational one to the ``operational" $\mathcal{O}_{1}$-basis (see Figure~\ref{fig:basis_change_gates}). Then CNOT gates are applied between the qubits $q_{1}, q_{2}, q_{3}$ and a delegation qubit $d_{1}$ to record the state onto that new qubit (see Figure~\ref{fig:circuit}). The inverse gates are applied to the $q_{i}$ to revert them back to the original state. Finally, this process is repeated for operators $\mathcal{O}_{i}$ and delegation qubits $d_{i}$ for $i=2,3$. The states of the delegation qubits are then measured to record the measurement outcomes of the operators $\mathcal{O}_{1}, \mathcal{O}_{2}, \mathcal{O}_{3}$.
\begin{figure}[ht]
\centering
    \begin{tabular*}{0.35\textwidth}{c|c}
        Operator & Gates \\ \hline 
        $I$ &$\cdot$ \\
        $X$ & \Qcircuit @C=1em @R=.7em {
        \lstick{} & \gate{H} & \qw
        } \\
        $Y$ & \Qcircuit @C=1em @R=.7em {
        \lstick{} & \gate{S^{\dag}} & \gate{H} & \qw
        } \\
        $Z$ & \Qcircuit @C=1em @R=.7em {
        \lstick{} & \qw & \qw &\qw
        }
    \end{tabular*}
    \caption{Basis change gates to convert from the computational basis to the ``operational" one, for a given context operator. For the identity $I$, no basis change gates nor CNOT gates are applied.}
    \label{fig:basis_change_gates}
\end{figure}

For example, in measuring $\mathcal{O}_{1} = XIY$, the gate $H$ is applied to $q_{1}$, no gate to $q_{2}$, and $S^{\dag} H$ to $q_{3}$. Then CNOTs are applied between $q_{1}$ and $d_{1}$, and $q_{3}$ and $d_{1}$, with $d_{1}$ acting as the target in each case. Inverse gates $H$, $HS$ are then applied to $q_{1}, q_{3}$ respectively, before measurement taken on $d_{1}$.

We have implemented the above circuitry for two contextual geometries: the elliptic quadric $\mathcal{Q}^{-}(5,2)$ and the complement $\overline{\mathcal{H}_{S}}$ of the skew-embedded hexagon. The measured contextuality $\chi$ for each is given in 
Table~\ref{tab:contextuality_exp}, 
compared with the QM and HV upper bounds. In each case, individual contexts were measured over $\numprint{8192}$ shots on the ``Lagos" IBM NISQ backend and also on a noiseless simulated quantum backend. When run on the simulated backend, the results give $\chi_{sim} = N$ as expected from~\eqref{eq:cabello_ineq}, however the inherent noise in the ``Lagos" backend reduces the measured value of $\chi_{NISQ}$ to somewhere still above the HV bound.

\begin{table}[pth!]
    \centering
    \begin{tabular}{cc|cc|cc}
        Geometry & $d$ & $N$ & $N-2d$ & $\chi_{sim}$ & $\chi_{NISQ}$ \\ \hline
        $\mathcal{Q}^{-}(5,2)$ & 9 & 45 & 27 & 45 & 27.86328 \\
        $\overline{\mathcal{H}_{S}}$ & 24 & 252 & 204 & 252 & 212.53735
    \end{tabular}
    \caption{Results for experimental tests of Cabello contextuality measure $\chi$ run on both a noiseless simulator and the IBM ``Lagos" backend. In both geometries $\mathcal{Q}^{-}(5,2)$ and $\overline{\mathcal{H}_{S}}$, the measured contextuality $\chi_{NISQ}$ is greater than the upper bound predicted by a non-contextual hidden variable model HV.\label{tab:contextuality_exp}}
\end{table}

Note that in both cases our experiments reveal the contextual nature of the configuration.
Our procedure to test Cabello's inequalities follow and slightly improves the experiments of~\cites{laghaout2022demonstration,holweck2021testing}. In particular, this time we delegated the measurements corresponding to the three observables of a context to three different delegation qubits (instead of a single qubit used in the previous cited work). Although this new procedure reduces the gap between the experimental value and the classical bound, it has the advantage of collecting the result for each of the three observables independently.

\section{Conclusion}\label{sec:conclusion}
This paper provides substantial insights into the role played by the smallest split Cayley hexagon $\mathcal{H}$ in its classical embeddings into $\mathcal{W}(5,2)$  in observable-based three-qubit contextuality, as first elaborated by three of us in~\cite{holweck2022three}. We have demonstrated that classically-embedded copies of $\mathcal{H}$ fully encode the basic information about distinguished contextual configurations living in $\mathcal{W}(5,2)$, namely doilies, both types of quadrics, complements of skew-embedded $\mathcal{H}$'s as well as the configuration formed by all the 315 lines of the space. Given such a configuration, one can always find some classically-embedded copy of $\mathcal{H}$ that shares with this configuration exactly those contexts that are unsatisfiable by an NCHV model! It is truly amazing to realize that the three-qubit $\mathcal{W}(5,2)$ is endowed with the distinguished subgeometry, $\mathcal{H}$,  that -- although being itself non-contextual -- is able to single out from a large variety of contextual configurations exactly those parts that `responsible' for their contextual behavior.  

Interestingly, we have already at hand some hints that something similar is taking place in the next case in the hierarchy -- the four-qubit $\mathcal{W}(7,2)$~\cite{muller-4to6qubit}. 
By making use of the Lagrangian Grassmannian mapping of type  $LGr(3,6)$ that sends planes of $\mathcal{W}(5,2)$ into points of a hyperbolic quadric $\mathcal{Q}^{+}(7,2)$ of $\mathcal{W}(7,2)$ we already found on this quadric a particular configuration that can be regarded as a four-qubit analog of a three-qubit Heawood-graph-underpinned
configuration described in Section~\ref{sec:class3qubit} (cf. Figure~\ref{hex-hyp}). This particular configuration contains 135 points and 315 lines, with seven lines through a point and three points on  a line, and -- being isomorphic to the dual polar space $\mathcal{DW}(5,2)$ -- has the desired property that it shares with {\it each} of the 120 $\mathcal{W}(5,2)$'s located on the $\mathcal{Q}^{+}(7,2)$ a copy of $\mathcal{H}$, the latter being indeed embedded classically into the corresponding $\mathcal{W}(5,2)$; moreover, it also contains 36 (one per each hyperbolic quadric of $\mathcal{W}(5,2)$ as dictated by $LGr(3,6)$-correspondence) copies of the point-plane incidence graph of PG$(3,2)$. What remains to be checked is whether there exists a four-qubit analog of a `classical' $\mathcal{H}$ of $\mathcal{W}(5,2)$, that is a configuration that picks up from each $\mathcal{Q}^{+}(7,2)$ of $\mathcal{W}(7,2)$ a configuration isomorphic to our $(135_7, 315_3)$-one. An affirmative answer to this computationally rather challenging task would mean, among other things, that the degree of contextuality of a four-qubit hyperbolic quadric is 315 and that of the whole  $\mathcal{W}(7,2)$ amounts to \numprint{1575}.

Finally, as suggested by one of the reviewers, it would be worth having a closer look if there is something deeper behind a seemingly formal analogy between the {\it two} inequivalent embeddings of the smallest split Cayley hexagon into the three-qubit $\mathcal{W}(5,2)$ and {\it two} inequivalent kinds of genuine tripartite entanglement, represented by the $GHZ$ state and the $W$ state~\cites{dvc00,cabad2002}. 
As it is well known, the entanglement of the $GHZ$ state~\cite{ghz90} disappears if any of the three qubits is traced over, whereas the entanglement in the $W$ state survives the loss of any of the three qubits; in other words, the entanglement in the $W$ state is robust
against the loss of one qubit, while the $GHZ$ state is reduced to a product of two qubits. Let us perform a similar trace-out-a-single-qubit procedure on the 63 observables of a three-qubit hexagon and see
which out of the 63 lines of such a hexagon retain their totally-isotropic character, i.\,e. reduce to two-qubit lines. A brief
inspection shows that in the case of a classically-embedded hexagon these two-qubit lines cover all the points except for those particular three ones that feature $I$ on both non-traced-out positions. A skew-embedded hexagon contains, however, additional points
that are left out, forming several different patterns. One of them features six additional points that lie in triples on two disjoint lines; moreover, and quite remarkably, these two lines define a unique grid whose
third line that is disjoint from the two is nothing but the axis of the corresponding hexagon.
So, like in the entanglement case, this observable-related tracing-out procedure leads to different outcomes in dependence on the way a hexagon is embedded into $\mathcal{W}(5,2)$. The very fact that the axis of a skew-embedded hexagon emerges also in this analogy is of particular interest. For example, it may lend itself as a sort of guiding principle when addressing the above-discussed four-qubit case: one can assume that some of the embeddings of a yet-to-be-discovered geometric configuration ruling four-qubit contextuality could well feature a distinguished linear subspace having one more dimension -- i.\,e. a {\it plane}. Moreover, given the fact that there are as many as nine inequivalent forms of four-qubit entanglement~\cite{verst-4qubits}, our ruling configuration should have eight more companions having the same underlying geometry but being embedded differently into $\mathcal{W}(7,2)$.
And this is certainly a topic that deserves to be properly treated in a separate paper.

\section*{Acknowledgments}
\vspace*{-0.3cm}
This work was supported in part by the Slovak VEGA Grant Agency, project number 2/0043/24,  by the PEPR integrated project EPiQ ANR-22-PETQ-0007 part of Plan France 2030 and by the project TACTICQ of the EIPHI Graduate School (contract ANR-17-EURE-0002). We acknowledge the use of the IBM Quantum Experience
and the IBMQ-research program. 
The views expressed are those of the authors and do not
reflect the official policy or position of IBM or the IBM Quantum Experience team. One
would like to thank the developers of the open-source framework Qiskit. We also thank 
Mr. Zsolt Szab\'o (Robert Bosch Kft, Budapest, Hungary) for the help with Figure~\ref{hwd}, Dr. Petr Pracna (Technology Centre ASCR, Prague, Czech Republic) for his invaluable help in preparing Figures~\ref{class-1st-kind} to~\ref{hex-hyp} as well as Dr. P\'eter L\'evay (Budapest University of Technology, Budapest, Hungary) for sharing his view on the action of $Sp(6,2)$ on $\mathcal{W}(5,2)$. 
Last but not least, we are also grateful to the reviewers for their useful remarks/suggestions.

\section*{Data availability}
All codes and results of our quantum experiments are openly available at:
\url{https://quantcert.github.io/hexagon}. The computations of degrees of contextuality are done with the software Qontextium, openly available at \url{https://quantcert.github.io/contextualityDegree/}.


\begin{thebibliography}{99}
\bibitem{ibmq}
IBM Quantum Experience. \url{https://quantum-computing.ibm.com/}
\bibitem{alla16}
D. Alsina and J. I. Latorre, Experimental test of Mermin inequalities on a five-qubit quantum
computer, Physical Review A 94 (2016), 012314.
\doi{10.1103/PhysRevA.94.012314}
\bibitem{appl05}
D. M. Appleby, The Kochen-Specker theorem, Studies in History and Philosophy of Modern Physics
36 (2005), 1--28. \doi{10.1016/j.shpsb.2004.05.003}
\bibitem{bell66}
J. S. Bell, On the problem of hidden variables in quantum mechanics, Reviews of Modern Physics
38 (1966), 447--452. \doi{10.1103/RevModPhys.38.447}
\bibitem{brav23}
J. Brody and R. Avram, Testing a Bell inequality with a remote quantum processor, Physics Teacher
61 (2023), 218--221. \doi{10.1119/5.0069073}
\bibitem{budroni2022kochen}
C. Budroni, A. Cabello, O. G\"uhne, M. Kleinmann, and J.-{\AA}. Larsson, Kochen-Specker contextuality,
Reviews of Modern Physics 94 (2022), 045007. \doi{10.1103/RevModPhys.94.045007}
\bibitem{Cab10}
A. Cabello, Proposed test of macroscopic quantum contextuality, Physical Review A 82 (2010),
032110. \doi{10.1103/PhysRevA.82.032110}
\bibitem{cabad2002}
A. Cabello, Bell’s theorem with and without inequalities for the three-qubit Greenberger-Horne-Zeilinger 
and W states, Phys. Rev. A 65 (2002), 032108. \doi{10.1103/PhysRevA.65.032108}
\bibitem{cool}
K. Coolsaet, The smallest split Cayley hexagon has two symplectic embeddings, Finite Fields and
Their Applications 16 (2010), 380--384. \doi{10.1016/j.ffa.2010.06.003}
\bibitem{DHGMS22}
H. de Boutray, F. Holweck, A. Giorgetti, P.-A. Masson, and M. Saniga, Contextuality degree of
quadrics in multi-qubit symplectic polar spaces, Journal of Physics A: Mathematical and Theoretical
55 (2022), 475301. \doi{10.1088/1751-8121/aca36f}
\bibitem{dvc00}
W. D\"ur, G. Vidal, and J. I. Cirac, Three qubits can be entangled in two inequivalent ways, Phys.
Rev. A 62 (2000), 062314. \doi{10.1103/PhysRevA.62.062314}
\bibitem{fano}
G. Fano, Sui postulati fondamentali della geometria proiettiva in uno spazio lineare a un numero
qualunque di dimensioni, Giornale di Matematiche 30 (1892), 106--132.
\url{http://www.bdim.eu/item?fmt=pdf&id=GM_Fano_1892_1}
\bibitem{ghz90}
D. M. Greenberger, M. A. Horne, A. Shimony, and A. Zeilinger, Bell’s theorem without inequalities,
American Journal of Physics 58 (1990), 1131–1143. \doi{10.1119/1.16243}
\bibitem{havlicek2009factor}
H. Havlicek, B. Odehnal, and M. Saniga, Factor-group-generated polar spaces and (multi)-qudits,
SIGMA. Symmetry, Integrability and Geometry: Methods and Applications 5 (2009), 096. 
\doi{10.3842/SIGMA.2009.096}
\bibitem{hs08}
H. Havlicek and M. Saniga, Projective ring line of an arbitrary single qudit, Journal of Physics
A: Mathematical and Theoretical 41 (2008), 225305. \doi{10.1088/1751-8113/41/1/015302}
\bibitem{heaw}
P. J. Heawood, Map colouring theorems, Quarterly Journal of Mathematics -- Oxford Series 24
(1890), 322–339.
\bibitem{holweck2021testing}
F. Holweck, Testing quantum contextuality of binary symplectic polar spaces on a noisy intermediate
scale quantum computer, Quantum Information Processing 20 (2021), 247. 
\doi{10.1007/s11128-021-03188-9}
\bibitem{holweck2022three}
F. Holweck, H. de Boutray, and M. Saniga, Three-qubit-embedded split Cayley hexagon is contextuality
sensitive, Scientific Reports 12 (2022), 8915. \doi{10.1038/s41598-022-13079-3}
\bibitem{HS17}
F. Holweck and M. Saniga, Contextuality with a small number of observables, International Journal
of Quantum Information 15 (2017), 1750026. \doi{10.1142/S0219749917500265}
\bibitem{Kelleher_2_qubit_games}
C. Kelleher, M. Roomy, and F. Holweck, Implementing 2-qubit pseudo-telepathy games on noisy
intermediate scale quantum computers, Quantum Information Processing 23 (2024), 187. 
\doi{10.1007/s11128-024-04386-x}
\bibitem{kosp67}
S. Kochen and E. Specker, The problem of hidden variables in quantum mechanics, Journal of
Mathematics and Mechanics 17 (1967), 59--87. \url{www.jstor.org/stable/24902153}, Ernst Specker Selecta (1990), 235--263. \doi{10.1007/978-3-0348-9259-9_21}
\bibitem{laghaout2022demonstration}
A. Laghaout, A. Dikme, N. Reichel, and G. Bj\" ork, A demonstration of contextuality using quantum
computers, European Journal of Physics 43 (2022), 055401. 
\doi{10.1088/1361-6404/ac79e0}
\bibitem{levay2017magic}
P. L\' evay, F. Holweck, and M. Saniga, Magic three-qubit Veldkamp line: A finite geometric underpinning
for form theories of gravity and black hole entropy, Physical Review D 96 (2017), 026018.
\doi{10.1103/PhysRevD.96.026018}
\bibitem{lps13}
P. L\' evay, M. Planat, and M. Saniga, Grassmannian connection between three- and four-qubit observables,
Mermin’s contextuality and black holes, Journal of High Energy Physics 9 (2013), 037.
\doi{10.1007/JHEP09(2013)037}
\bibitem{lsv}
P. L\' evay, M. Saniga, and P. Vrana, Three-qubit operators, the split Cayley hexagon of order two
and black holes, Physical Review D 78 (2008), 124022. \doi{10.1103/PhysRevD.78.124022}
\bibitem{ls17}
P. L\' evay and Z. Szab\'o, Mermin pentagrams arising from Veldkamp lines for three qubits, Journal
of Physics A: Mathematical and Theoretical 50 (2017), 095201. 
\doi{10.1088/1751-8121/aa56aa}
\bibitem{mermin}
N. D. Mermin, Hidden variables and the two theorems of John Bell, Reviews of Modern Physics
65 (1993), 803--815. \doi{10.1103/RevModPhys.65.803}
\bibitem{mg23}
A. Muller and A. Giorgetti, An abstract structure that determines the degree of contextuality
of observable-based Kochen-Specker proofs, 2024. arXiv preprint \url{www.arxiv.org/abs/2410.14463}, 
submitted for publication.
\bibitem{muller2023new}
A. Muller, M. Saniga, A. Giorgetti, H. de Boutray, and F. Holweck, New and improved bounds on
the contextuality degree of multi-qubit configurations, Mathematical Structures in Computer Science
34 (2024), 322–343. \doi{10.1017/S0960129524000057}
\bibitem{muller-4to6qubit}
A. Muller, M. Saniga, A. Giorgetti, F. Holweck, and C. Kelleher, A new heuristic approach for
contextuality degree estimates and its four- to six-qubit portrayals, 2024. arXiv preprint 
\url{www.arxiv.org/abs/2407.02928}, submitted for publication.
\bibitem{peres1991two}
A. Peres, Two simple proofs of the Kochen-Specker theorem, Journal of Physics A: Mathematical
and General 24 (1991), L175--L178. \doi{10.1088/0305-4470/24/4/003}
\bibitem{psh}
M. Planat, M. Saniga, and F. Holweck, Distinguished three-qubit `magicity' via automorphisms of
the split Cayley hexagon, Quantum Information Processing 12 (2013), 2535--2549. 
\doi{10.1007/s11128-013-0547-3}
\bibitem{psm}
B. Polster, A. E Schroth, and H. Van Maldeghem, Generalized flatland, The Mathematical Intelligencer
23 (2001), 33--47. \doi{10.1007/BF03024601}
\bibitem{sbhg21}
M. Saniga, H. de Boutray, F. Holweck, and A. Giorgetti, Taxonomy of polar subspaces of multiqubit
symplectic polar spaces of small rank, Mathematics 9 (2021), 2272. \doi{10.3390/math9182272}
\bibitem{slp12}
M. Saniga, P. L\' evay, and P. Pracna, Charting the real four-qubit Pauli group via ovoids of a
hyperbolic quadric of PG(7,2), Journal of Physics A: Mathematical and Theoretical 45 (2012),
295304. \doi{10.1088/1751-8113/45/29/295304}
\bibitem{sp06}
M. Saniga and M. Planat, Hjelmslev geometry of mutually unbiased bases, Journal of Physics A:
Mathematical and General 39 (2006), 435–440. \doi{10.1088/0305-4470/39/2/013}
\bibitem{saniga2007multiple}
M. Saniga and M. Planat, Multiple qubits as symplectic polar spaces of order two, Advanced Studies in Theoretical
Physics 1 (2007), 1--4. \url{www.arxiv.org/abs/quant-ph/0612179}
\bibitem{sppl}
M. Saniga, M. Planat, P. Pracna, and P. L\' evay, `Magic' configurations of three-qubit observables and
geometric hyperplanes of the smallest split Cayley hexagon, SIGMA. Symmetry, Integrability and
Geometry: Methods and Applications 8 (2012), 083. \doi{10.3842/SIGMA.2012.083}
\bibitem{schr}
A. E. Schroth, How to draw a hexagon, Discrete Applied Mathematics 199 (1999), 161--171. 
\doi{10.1016/S0012-365X(98)00294-5}
\bibitem{speck60}
E. Specker, Die Logik nicht gleichzeitig entscheidbarer Aussagen, Dialectica 14 (1960), 239--246.
\doi{10.1111/j.1746-8361.1960.tb00422.x}
\bibitem{koen09}
K. Thas, The geometry of generalized Pauli operators of N-qudit Hilbert space, and an application
to MUBs, EPL (Europhysics Letters) 86 (2009), 60005. \doi{10.1209/0295-5075/86/60005}
\bibitem{tran22}
D. M. Tran, D. V. Nguyen, B. H. Le, and H. Q. Nguyen, Experimenting quantum phenomena on
NISQ computers using high level quantum programming, EPJ Quantum Technology 9 (2022), 6.
\doi{10.1140/epjqt/s40507-022-00126-1}
\bibitem{mal}
H. Van Maldeghem, Generalized polygons, Springer Science $\&$ Business Media, 2012. 
\doi{10.1007/978-3-0348-0271-0}
\bibitem{verst-4qubits}
F. Verstraete, J. Dehaene, B. De Moor, and H. Verschelde, Four qubits can be entangled in nine
different ways, Phys. Rev. A 65 (2002), 052112. \doi{10.1103/PhysRevA.65.052112}
\bibitem{VL10}
P. Vrana and P. L\' evay, The Veldkamp space of multiple qubits, Journal of Physics A: Mathematical
and Theoretical 43 (2010), 125303. \doi{10.1088/1751-8113/43/12/125303}
\bibitem{wae14}
M. Waegell, Primitive nonclassical structures of the N-qubit Pauli group, Physical Review A 89
(2014), 012321. \doi{10.1103/PhysRevA.89.012321}
\bibitem{waeara13}
M. Waegell and P. K. Aravind, Proofs of the Kochen–Specker theorem based on the N-qubit Pauli
group, Physical Review A 88 (2013), 012102. \doi{10.1103/PhysRevA.88.012102}
\bibitem{xu_experimental_2022}
J.-M. Xu, Y.-Z. Zhen, Y.-X. Yang, Z.-M. Cheng, Z.-C. Ren, K. Chen, X.-L. Wang, and H.-T.
Wang, Experimental Demonstration of Quantum Pseudotelepathy, Physical Review Letters 129
(2022), 050402. \doi{10.1103/PhysRevLett.129.050402}
\end{thebibliography}
\end{document}